\documentclass[aps,prb,twocolumn,amsmath,longbibliography,amssymb,superscriptaddress]{revtex4-2}
\usepackage{bm}
\usepackage{color}
\usepackage{graphicx}
\usepackage{bbold}
\newcommand{\bq}{\begin{equation}}
\newcommand{\ee}{\end{equation}}
\DeclareMathOperator{\sign}{sign}

\begin{document}

\title{Collective spin oscillations in a magnetized graphene sheet}
\author{M. Agarwal}
\affiliation{Department of Physics and Astronomy, University of Utah, Salt Lake City, Utah 84112, USA}

\author{O. A. Starykh}
\affiliation{Department of Physics and Astronomy, University of Utah, Salt Lake City, Utah 84112, USA}

\author{D. A. Pesin}
\affiliation{Department of Physics, University of Virginia, Charlottesville, Virginia 22904, USA}

\author{E. G. Mishchenko}
\affiliation{Department of Physics and Astronomy, University of Utah, Salt Lake City, Utah 84112, USA}

\begin{abstract}
We investigate collective spin excitations of graphene electrons with short-ranged interactions and subject to the external Zeeman magnetic field.  We find that in addition to the familiar Silin spin wave, a collective spin-flip excitation that reduces to the uniform precession when the wave’s momentum approaches zero, the magnetized graphene supports another collective mode visible in the transverse spin susceptibility: a collective spin-current mode. Unlike the Silin wave, this mode is not dictated by the spin-rotational symmetry but rather owns its existence to the pseudo-spin structure of the graphene lattice. We find the new collective excitation to become sharply defined in a finite interval of wave's momenta, the range of which is determined by the interaction and the magnetization. 
\end{abstract}

\maketitle
\section{Introduction}

The spin response of paramagnetic metals subjected to the external magnetic field represents one of the storied research directions in condensed matter physics. These studies, initiated by Silin in 1958 within the framework of the Landau Fermi-liquid theory \cite{Silin1958,Silin1959}, produced a detailed theoretical understanding of collective spin oscillations of itinerant interacting fermion systems  
\cite{Platzman1967,Leggett1970,vanLoon2023} whose key features were successfully tested by experiments \cite{Schultz1967}. The principal prediction of these studies, which focus on single-band conductors with parabolic electron dispersion, is the emergence of the transverse collective spin mode below the particle-hole continuum. This downward dispersing collective mode originates from the Zeeman frequency at zero wave vector where it represents the collective precession of the total magnetic moment as required by the Larmor theorem. The very existence of this mode as a collective excitation of the Fermi-liquid is based on the interacting nature of the latter and allows one to qualitatively distinguish the interacting Fermi liquid ground state from the non-interacting Fermi gas state.

Our paper is devoted to the calculation of the dynamic spin susceptibility of the Zeeman-field-magnetized monolayer graphene at the charge-neutrality point (filled valence band and empty conduction band when the Zeeman field is turned off). 
This limit is opposite to the standard case of the single-band conductor described above, in which the Zeeman energy is several orders of magnitude smaller than the Fermi energy. 

To the best of our knowledge, the spin sector of graphene has been relatively underexplored. 
Collective spin excitations of non-magnetized graphene were studied early on \cite{Baskaran2002} and claimed the existence of the coherent spin-1 branch below the particle-hole continuum. However, this finding was subsequently refuted \cite{anti-Baskaran2004}.
Two recent papers \cite{raines2021,raines2022} have studied charge-neutral collective modes of monolayer graphene, including the case with finite $\omega_B$, but focused on the inter-valley excitations in the electron-doped regime when $E_F \gg \omega_B$. A Silin-like spin mode of the doped graphene was also studied via a time-dependent
spin-density-functional response theory \cite{Anderson2021}.

The equilibrium properties of a single graphene sheet partially polarized by an in-plane magnetic field -- the setup considered in this work -- have been theoretically analyzed in Ref.~\cite{aleiner2007}. It was found that at temperatures significantly below the Zeeman splitting scale $\omega_B$, the long-range Coulomb interaction drives graphene into the correlated exciton insulator state. Our investigation of the dynamic spin response at the energies of order $\omega_B$ is complementary to such low energy/temperature range where the exciton insulator is predicted to occur.

The limit that we consider corresponds to the interesting situation of Zeeman energy $\omega_B$ exceeding the Fermi energy $E_F$ (which is exactly zero at the charge-neutrality point). We show below that it leads to the appearance of a new collective spin mode associated with oscillations of the spin current  that is found to co-exist with the usual transverse Larmor-Silin spin wave.

The rest of the paper is organized as follows: In Section~\ref{sec:model} we introduce the model of a magnetized graphene sheet, and discuss the origin of spin collective modes qualitatively. In Section~\ref{sec:interaction} we study the interacting Green's functions of the system, present the expression for the relevant polarization functions, calculated in the Appendix, and use them to solve the Bethe-Salpeter equation for the vertex function. In Section~\ref{sec:modes} we determine the spectra of collective modes. Finally, in Section~\ref{sec:discussion} we summarize and discuss the quantitative results of this work.

\section{The model and qualitative picture of collective modes}\label{sec:model}

In this Section we present a model of undoped interacting graphene sheet in an external Zeeman field, and discuss the qualitative picture of spin and spin-pseudospin collective modes in this system. 

\subsection{Model of undoped magnetized graphene}

We consider two-dimensional Dirac fermions in the presence of an applied external magnetic field. We direct the $z$-axis in the spin space along the $x$-axis in the real space, such that in the real-space coordinates the magnetic field is given by ${\bf B} = (-B_0,0,0)$, $B_0>0$. Such magnetic field is purely Zeeman and does not affect the orbital motion of electrons. The negative sign of the magnetic field is chosen for convenience, to compensate for
the negative sign of the electron charge. This means 
that the energy of spin-up electrons is decreased by the magnetic field whereas the energy of spin-down electrons is increased by it.  Since we do not consider effects of spin-orbit or dipolar coupling, the fact that the $z$-axis in the spin space points along the $x$-axis in the real space does not lead to any difficulties. 

Associated with the Zeeman field is the  spin splitting of the energy bands, $\omega_B = g \mu_B B_0/2$, where $g$ is the electron g-factor and $\mu_B$ is the Bohr magneton. Below the two real spin states are denoted with $\uparrow,\downarrow$; the operators acting on these states are denoted with Pauli matrices $\tau_0,\bm \tau$. The two sublattice indices, which we will also refer to as pseudospin, are denoted with $A,B$; the operators acting on the pseudospin states are denoted with Pauli matrices $\sigma_0,\bm \sigma$. The Hamiltonian for a Dirac point is written in the Kronecker product space of spin and pseudospin. We will use letters from the beginning of the Greek alphabet to label states in the four-dimensional Kronecker product space spanned by $(\uparrow A,\uparrow B,\downarrow A,\downarrow B)$. The corresponding field operators, $\hat\Psi_{\alpha}(\bm r)$, and the annihilation operators for the electrons in momentum space, $\hat c_{\bm k \alpha}$, are related by $\hat\Psi_{\alpha}(\bm r)=\sum_{\bm k}e^{i\bm k \bm r}\hat c_{\bm k\alpha}$, with the normalization area set to unity.
We will also work in units in which the electron Fermi velocity and the Planck constant are set to unity.

The non-interacting part of the Hamiltonian for a given Dirac point can be written as
\begin{align}
\label{non-interacting hamiltonian}
\hat H & = \sum_{{\bf k}}  \hat c^\dagger_{\bf k,\alpha} \left[\tau_0 \otimes  {\bm \sigma} \cdot {\bf k} -\omega_B \tau_z \otimes \sigma_0 \right]_{\alpha\beta} \hat c_{\bf k,\beta}.
\end{align}
The Zeeman field modifies the energy bands, ${\cal E}_{\tau}^{s}({\bf k}) = -\tau \omega_{B} + s vk$, where $s=1$ corresponds to the upper Dirac cones (conduction bands for a given spin $\tau$), and $s=-1$ corresponds to the lower Dirac cones (valence bands for a given spin $\tau$). This is consistent  with the energy of  spin-up electrons being decreased by the magnetic field, and the energy of spin-down electrons being increased by it, as shown in Fig.~\ref{fig:occupiedstates}.  

To study collective modes of the system we choose a simplified form of interaction:
\begin{equation}
\label{interacting hamiltonian}
\hat H_{int} =  \frac{1}{2} \sum_{{\bf p,k,q}} U_q \hat c_{{\bf k+q},\alpha}^{\dagger}\hat c_{\bf{p-q},\beta}^{\dagger} \hat c_{{\bf p},\beta}\hat c_{{\bf k},\alpha}.
\end{equation}
The short-range part of such interaction Hamiltonian contains both the intra-sublattice Hubbard interaction and the inter-sublattice Coulomb repulsion. These types of interaction obviously should have different strengths, but in graphene they are only different by about a factor of two~\cite{katsnelson2011coulomb}. Hence for simplicity we consider the interaction of higher spin-sublattice symmetry, Eq.~\eqref{interacting hamiltonian}. This spurious symmetry does not lead to any qualitative effects in the present context. The  important aspect of the chosen interaction Hamiltonian is that it conserves both real spin and pseudospin, as explained in the next section. 

\subsection{Qualitative picture of collective modes}

We focus on the neutral collective modes of the system, limiting ourselves to those involving collective transverse spin motions and excluding pure charge (plasmon) excitations. 

\begin{figure}
    \centering
    \includegraphics[width=3.5in]{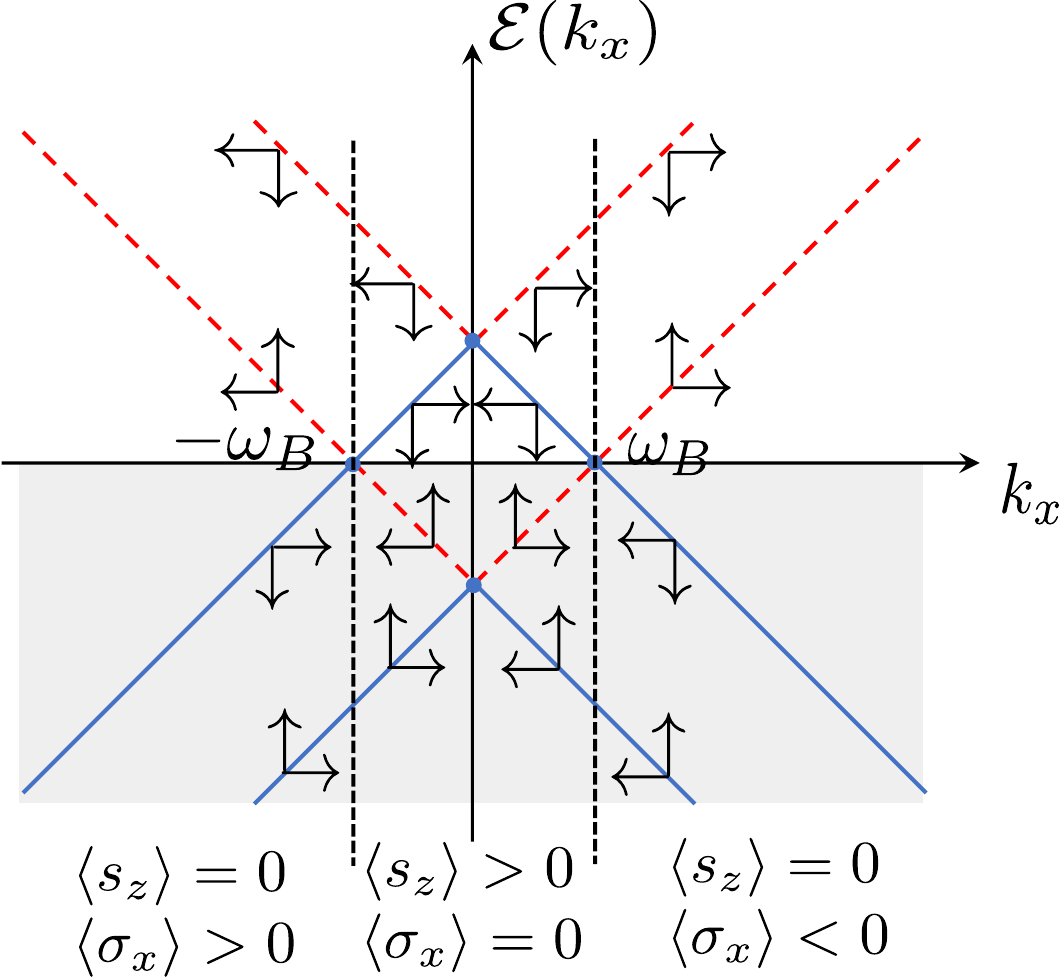}
    \caption{Scheme of the band structure and occupied states along the $k_x$ axis in the momentum space. Dashed red lines represent the conduction band states for spin-up (displaced downward in energy), and spin-down (displaced upward in energy) states. Solid blue lines are the valence bands for the two spin states. Vertical dotted lines separate regions of finite spin or pseudospin polarizations. The spin and pseudospin polarizations of the states are indicated with vertical and horizontal arrows positioned on the dispersion lines. }
    \label{fig:occupiedstates}
\end{figure}

First, consider the spin and pseudospin polarization patterns in the non-interacting ground state of the system, as shown in Fig.~\ref{fig:occupiedstates}. 
Since we are using units with Dirac speed set to unity, the scales for the relevant single-particle energies and momenta are all given by the Zeeman splitting of the bands, $\omega_B$. 

We note that the ground state is spin-polarized along the Zeeman field. At zero chemical potential, there is an excess $\delta n_{\uparrow}>0$ of spin-up electrons and an equal deficit $\delta n_{\downarrow}<0$ of spin-down electrons, where $\Delta n = \delta n_{\uparrow} - \delta n_{\downarrow} = 2 \delta n_{\uparrow} = \omega_{B}^2/2\pi$. The net spin polarization persists in the interacting ground state, and implies that small spin density deviations from the direction of the Zeeman field will precess around this field. The Larmor theorem, which stems from the commutativity of the interaction Hamiltonian with the total spin operator, guarantees that uniform spin precession occurs at the frequency of $2\omega_B$ even in an interacting system. The result of the celebrated works of Silin~\cite{Silin1958,Silin1959} is that this collective precession stays coherent even at finite wave vector, resulting in the existence of a collective wave of transverse spin fluctuations. 

We do find the Silin-like wave in magnetized graphene, see Eq.~\eqref{eq:silinmode}, and the text around it. It can be crudely thought of as collective oscillations of a transverse Zeeman field, $\langle\Psi^\dagger_{\alpha}(\bm r)\left[\tau_{x,y}\otimes\sigma_0\right]_{\alpha\beta}\Psi_{\beta}(\bm r)\rangle$, where summation over repeated indices is implied. In other words, it corresponds to fluctuations of interband coherence between occupied and unoccupied states with the opposite spin polarizations, but the same pseudospin polarization. In the long-wavelength limit, these states are confined to $|k|<\omega_B$, and the corresponding dispersion lines run parallel to each other in Fig.~\ref{fig:occupiedstates}, and are separated by energy of $2\omega_B$, which sets the frequency of the collective oscillation at zero wave vector.  

The view of the Silin mode as oscillations of interband coherences allows one to see what new features the Dirac spectrum of graphene brings into the problem, and ask new questions. The states with the same pseudospin, but opposite real spin orientations belong to pairs of  either conduction or valence bands, which got split by the Zeeman field. In this sense these coherences are the direct analogs of those appearing in a single-band Fermi liquid considered by Silin. In the case of graphene, with its Dirac band structure, one can then ask if there are oscillations of coherences between occupied and unoccupied states of \emph{both} opposite spin \emph{and} pseudospin polarizations. The corresponding dispersion lines have opposite slopes in Fig.~\ref{fig:occupiedstates}, and exist for all $k$'s. Such oscillations would correspond to a hybrid spin-pseudospin collective field, $\langle\Psi^\dagger_{\alpha}(\bm r)\left[\tau_{i}\otimes\sigma_j\right]_{\alpha\beta}\Psi_{\beta}(\bm r)\rangle$. We do find that such hybrid collective modes exist for $i,j=1,2$, and have a dispersion given by Eq.~\eqref{eq:hybridmode}. These modes are overdamped at small wave vector $q$, but do become well-defined for weak interactions before disappearing into a particle-hole continuum associated with spin-flip intraband transitions. Here and below `intraband' refers to transitions between spin-split bands of the same pseudospin polarization. These would be either degenerate conduction or valence bands without the Zeeman field. 

The existence of these hybrid spin-pseudospin modes is not due to any symmetry, since the total pseudospin operator does not commute with the noninteracting part of the Hamiltonian, hence there is no Larmor theorem for the precession of the corresponding observable, even though the interaction Hamiltonian conserves the pseudospin. 

\begin{figure}
    \centering
    \includegraphics[width=2.5in]{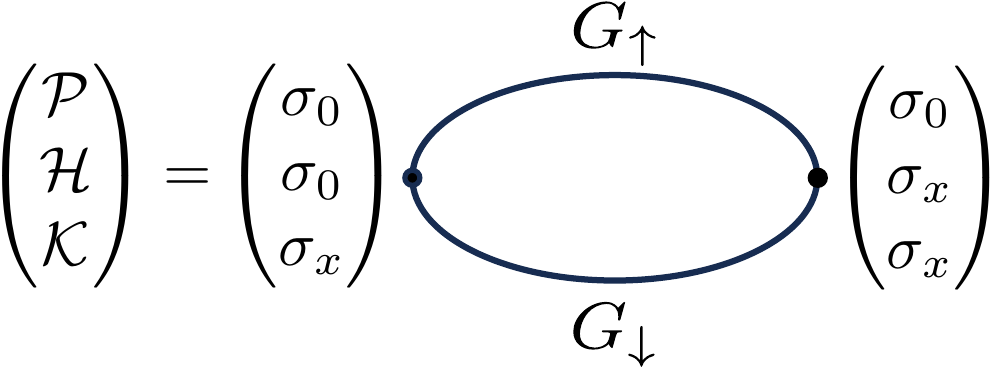}
    \caption{Polarization functions of electrons, which determine their response to a generalized Zeeman field with nontrivial matrix structure in the pseudospin space, see the discussion in Section~\ref{sec:model}. The solid lines are the single-particle Green's functions for spin-up and spin-down electrons, $G_{\uparrow,\downarrow}$, which are matrices in the pseudospin space. The quantities in parentheses should be read line by line: those on the left hand side of the graphical equation are the spin-spin, ${\cal P}$, spin-spin-current, ${\cal H}$, and spin-current-spin-current, $\cal K$, polarization functions, defined further in Section~\ref{subsec:polarizations}. Each line on the right hand side gives a pair of the corresponding vertices. }
    \label{fig:all-the-loops}
\end{figure}

We can proceed to study the collective excitation that involve transverse spin fluctuations in two ways. First, given the nature of the spin-pseudospin collective field described above, we can formally consider linear response of the system to a generalized Zeeman field, which has the matrix structure of $\tau_i\otimes\sigma_j$, with $i=1,2$. The $j=0$ component of such a field corresponds to the conventional Zeeman field. For noninteracting electrons, this linear response is determined by three types of polarization bubbles, schematically shown in Fig.~\ref{fig:all-the-loops}. The spectrum of the collective modes can then be inferred from the pole structure of the corresponding matrix generalized susceptibility, obtained from bubbles of Fig.~\ref{fig:all-the-loops} by dressing them with interaction lines. 

\begin{figure}
    \centering
    \includegraphics[width=2.0in]{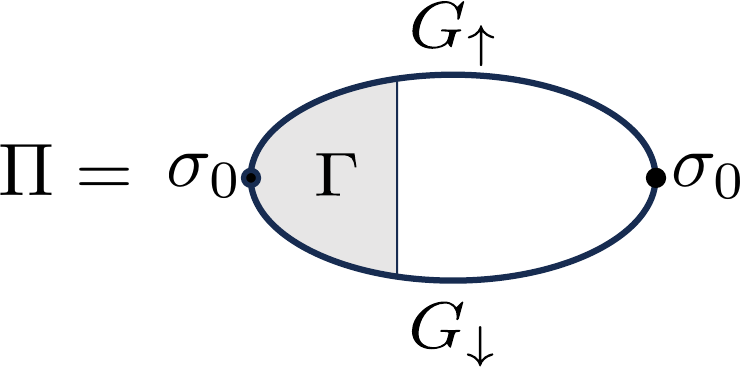}
    \caption{Polarization functions of electrons determining their transverse dynamical spin susceptibility. The shaded part is the vertex function, $\Gamma$, while the solid lines are the single-particle Green's functions of interacting electrons.}
    \label{fig:loop}
\end{figure}

Instead of the above approach, in this work we will interpret the hybrid spin-pseudospin modes as oscillations of the spin current, and argue that the corresponding collective mode spectrum can be obtained from the conventional transverse dynamical spin susceptibility, $\chi_\perp(\bm q,\omega)$. Indeed, for Hamiltonian~\eqref{non-interacting hamiltonian} the single-particle velocity operator coincides with the pseudospin polarization function, $\bm\sigma$, which means that the collective field $\langle\Psi^\dagger_{\alpha}(\bm r)\left[\tau_{i}\otimes\sigma_j\right]_{\alpha\beta}\Psi_{\beta}(\bm r)\rangle$ coincides with the un-symmetrized spin current operator. This means that nonuniform fluctuations of this field will give rise to net spin density fluctuations, as can be easily seen from the continuity equation for the spin density. We thus expect the spin-current collective modes to appear as resonances in the transverse dynamical spin susceptibility at $\bm q\neq 0$. It is also clear that the Silin mode, being a spin transverse spin wave, can also be studied via the same transverse spin susceptibility, as was done in Ref.~\cite{starykh2020silin} for a $U(1)$ spin liquid with a Fermi surface. Since $\chi_\perp(\bm q,\omega)$ is determined by a single electronic polarization loop shown in Fig.~\ref{fig:loop}, the spin-current-spin-current susceptibility, $\cal K$, and the spin-current-spin-density susceptibility, $\cal H$, shown in Fig.~\ref{fig:all-the-loops} will enter the calculations through the vertex function $\Gamma$ in Fig.~\ref{fig:loop}.
All of these polarization functions are calculated in the Appendix, while in the main part of the paper we use the results to study the collective modes in the system.

\section{Magnetized graphene with short-range interactions}\label{sec:interaction}

In this Section we present the zero-temperature Green's function and polarization functions for the model defined by Eqs.~\eqref{non-interacting hamiltonian} and~\eqref{interacting hamiltonian}, and use these results to solve the Bethe-Salpeter equation for the full spin-spin polarization function, diagrammatically defined in Fig.~\ref{fig:loop}. 

\subsection{Electron self-energy and the single-particle Green's functions}

We first determine the combined effect of the magnetic field and interaction on the electron self-energy, specializing to the case of a short-range interaction. 

\begin{figure}
 \includegraphics[width=7cm,height=7cm,keepaspectratio]{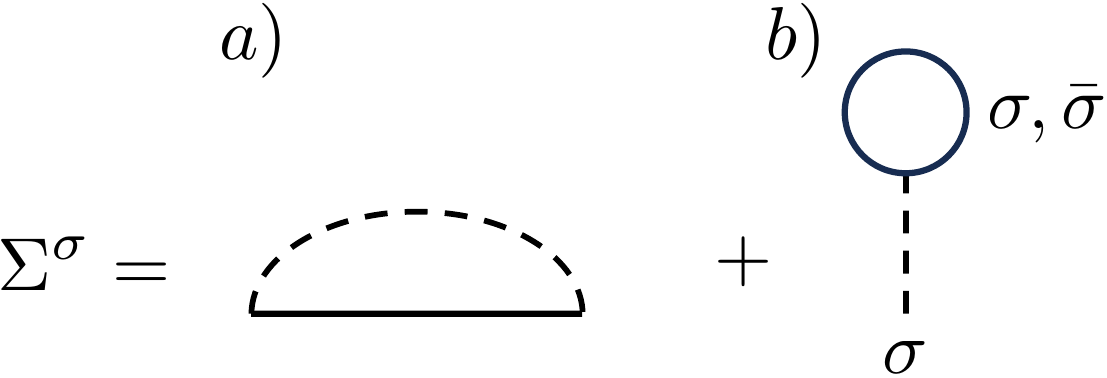}
\caption{\label{Fig. 3} First-order electron self-energy diagrams, a) exchange term, and, b) tadpole term.}
\end{figure}
The electron self-energy, to the first order in $U_q$, includes contribution from two diagrams, illustrated in Fig.~\ref{Fig. 3}a) and Fig.~\ref{Fig. 3}b), where solid lines indicate electron Green's functions, see Eq.~(\ref{Green's function}), and dashed lines correspond to the electron-electron interaction.
The exchange contribution, Fig.~\ref{Fig. 3}a), is
\begin{equation}
\label{Self-energy exchange}
{\textstyle \hat \Sigma_ {a}^{\tau}}({\bf p}) = i\sum_{{\bf k}}U_{|{\bf p}-{\bf k}|}  \int \frac{d\epsilon}{2\pi}  \hat {\cal G}_{\epsilon, {\bf k}}^{\tau} e^{i\epsilon \, 0+}.
\end{equation}
The factor $e^{i\epsilon \, 0+}$ selects the correct normal ordering of electron operators in the Green's function. 
The Hartree contribution, Fig.~(\ref{Fig. 3}b), is 
\begin{equation}
\label{Self-energy Hartree}
{\textstyle \hat \Sigma_ {b}^{\tau}}({\bf p})= -iU_{0}\sum_{{\bf k},\tau'} \int \frac{d\epsilon}{2\pi}  \hat {\cal G}_{\epsilon, {\bf k}}^{\tau'} e^{i\epsilon \, 0+}.
\end{equation}

For the short-range interaction, $U_q\equiv u$, the exchange contribution cancels the $\tau'=\tau$ part of the Hartree contribution, leaving only the term $\tau'=\bar \tau$, where the bar in $\bar \tau$ indicates the spin opposite to spin $\tau$:
\begin{align}
\label{Self-energy}
{\textstyle \hat \Sigma^{\tau}}&({\bf p}) = -iu \sum_{{\bf k}} \int \frac{d\epsilon}{2\pi}  \hat {\cal G}_{\epsilon, {\bf k}}^{\bar \tau} e^{i\epsilon \, 0+} \nonumber\\
								=&-\frac{iu }{2}\sum_{{\bf k},s = \pm1} \! \int \!\frac{d\epsilon}{2\pi} \frac{e^{i\epsilon \, 0+} }{\epsilon+\bar\tau \omega_{B} - sk +i\eta^{\bar\tau}_{s}(k)0^{+}} ,
\end{align}
The odd in ${\bf k}$ term $\hat\sigma_{\bf k}$ vanishes upon the ${\bf k}$-integration. The value $\eta^{\bar\tau}_s(k)$ is positive for empty states and negative for filled states. In particular, for spin-down electrons,
\begin{equation}
\eta^{\uparrow}_{s} (k) = 
					\begin{cases}
					\sign (k-\omega_B), & s = +1 \\
					-1, & s = -1
					\end{cases}
\end{equation}
and, spin-down electrons, 
\begin{equation}
\eta^{\downarrow}_{s}(k) = 
					\begin{cases}
					1, & s = +1 \\
					\sign(\omega_B - k), & s = -1
					\end{cases}
\end{equation}
Evaluating the energy integral by closing the integration contour in the upper half-plane and subtracting the infinite but field-independent contribution at $\omega_B=0$, we obtain the self-energy, 
${\textstyle \Sigma^{\tau}}({\bf p}) = u\delta n^{\bar\tau}/2$. For multiple Dirac points (e.g., $N=2$), the self-energy is further multiplied by the number of the Dirac points
${\textstyle \Sigma^{\tau}}({\bf p}) = N u\delta n^{\bar\tau}/2$.

With the self-energy included, the single-particle causal Green's function becomes 
\begin{equation}
\label{Modified Green's function}
\hat {\cal G}^{\tau}_{\epsilon,\bm{k}} = \frac{1}{2} \sum_{s = \pm1} \frac{1+s \hat \sigma_{\bm{k}}}{\epsilon- {\cal E}^{\tau}_{s}(k) - \Sigma^{\tau}+is0^+}.
\end{equation}

The expression for the Green's function shows that the self-energy effect for the short-range interaction reduces to the renormalization of the Zeeman energy in the single-particle Hamiltonian. For $N=2$ and $\delta n_{\uparrow}=-\delta n_{\downarrow}=\Delta n/2$ this renormalization is given by 
\begin{align}\label{eq:tildeomega}
    \omega_B\to \tilde\omega_B=\omega_B+u\Delta n/2. 
\end{align}

Consequently, the value of the net spin polarization, $\Delta n$, must be found from a self-consistent equation
\begin{align}
    \Delta n=\frac{\tilde\omega^2_B}{2\pi}.
\end{align}

\subsection{{Polarization functions}}\label{subsec:polarizations}

As explained in Section~\ref{sec:model}, we need three types of polarization functions to describe transverse spin response of the system. These are shown diagramatically in Fig.~\ref{fig:all-the-loops}. The analytic expressions for these functions are as follows. 

The transverse spin-spin polarization function, $\cal P$, as a function of external frequency $\omega$ and momentum ${\bf q}$, is given by 
\begin{equation}\label{Polarization a}
 {\mathcal P}(\omega,\bm{q}) = -i {N}  \text{Tr}_\sigma \sum_{\bm{k}} \int\frac{d\epsilon}{2\pi}  \hat {\cal G}^{\downarrow}_{\epsilon_{+},\, {\bf k}_+} \hat {\cal G}^{\uparrow}_{\epsilon_{-}, \,{\bf k}_-},
\end{equation}
where $N=2$ is the number of Dirac cones and $\epsilon_{\pm} =\epsilon \pm \omega/2$,  ${\bf k}_\pm ={\bf k}\pm{\bf q}/2$.

The spin-current-spin-current polarization function is given by 
\begin{equation}
\label{Current-current correlation a}
{\cal K}
(\omega,\bm{q}) = -i {N}  \text{Tr}_\sigma \sum_{\bm{k}} \int\frac{d\epsilon}{2\pi} \hat {\cal G}^{\downarrow}_{\epsilon_{+}, {\bf k}_+}\sigma_x \hat {\cal G}^{\uparrow}_{\epsilon_{-},{\bf k}_-}\sigma_x.
 \end{equation}

Finally, the mixed spin-spin-current polarization function is 
\begin{equation}
{\cal H}(\omega,\bm{q}) = -i \frac{N}{2}  \text{Tr}_\sigma \sum_{\bm{k}} \int\frac{d\epsilon}{2\pi}  G^{\downarrow}_{\epsilon_{+}, {\bf k}_+}\sigma_x \hat G^{\uparrow}_{\epsilon_{-} ,{\bf k}_-}.
 \end{equation}

The Green's functions entering the expressions for the polarization functions include the self-energy correction, Eq.~\eqref{Modified Green's function}. The full calculation of these functions is given in the Appendix. Here we quote the final results, needed for the analysis of the collective modes below. 

It is convenient to present the polarization functions as functions of the shifted value of the frequency $\omega$:
\begin{align}
    \Omega\equiv\omega - 2\omega_B-u\Delta n = \omega-2\tilde\omega_B,
\end{align}
where the definition of $\tilde\omega_B$ is the same as given in Eq.~\eqref{eq:tildeomega}.

We start with the spin-spin polarization, $ {\cal P}(\omega,q)$, see Appendix~\ref{appendix:spin-spin} for details.
\begin{figure}
    \centering
    \includegraphics[width=2.5in]{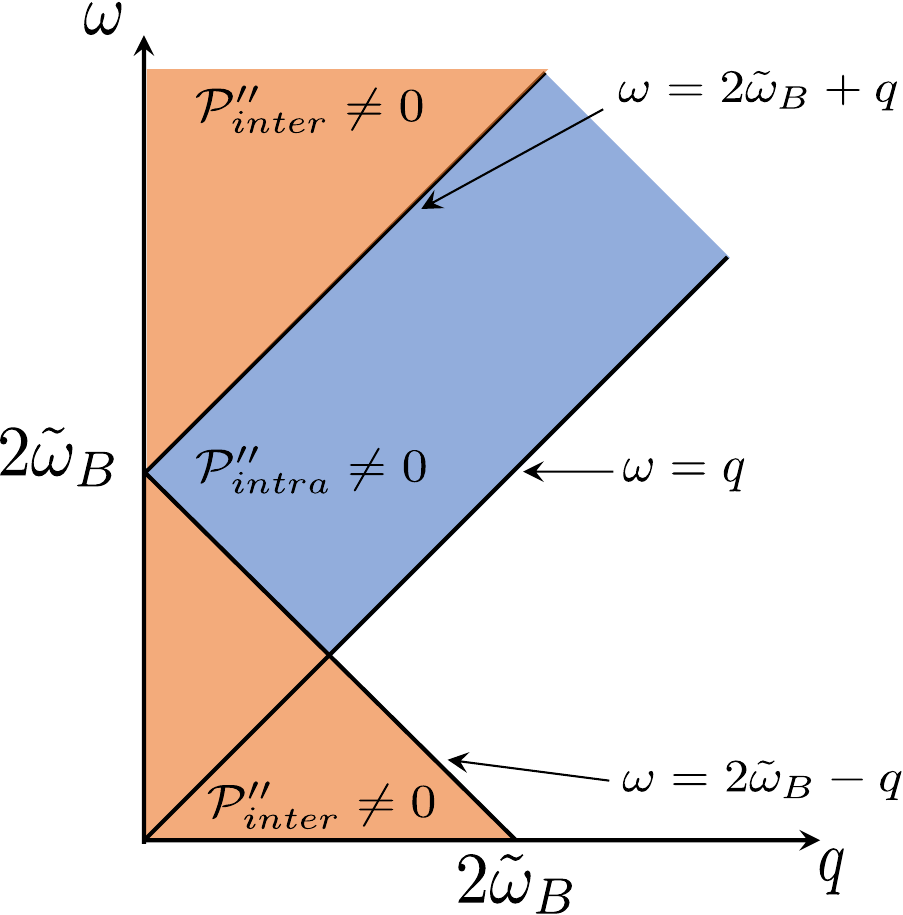}
    \caption{Shaded region shows where imaginary part of  polarization function $\cal P$ coming from intraband (blue-shaded) and interband (orange-shaded) transitions is non-zero.}
    \label{fig:p-hcontinuum}
\end{figure}
We present its form for $\Omega<0$, $|\Omega|>q$, and $\omega_{B} \gg q$, which is the relevant region for the determination of the collective mode spectrum:
\begin{equation}
\label{Non-interacting Polarization}
 {\mathcal P}(\omega,q) = -\frac{N \omega_{B}^2}{2\pi\sqrt{\Omega^2 - q^2}}
										-i\frac{N q^2}{16\sqrt{\Omega^2 - q^2}}.
\end{equation}

The structure of the imaginary part of $ {\cal P}(\omega,q)$ for general $\omega,q$, and the spin-flip particle-hole continuum is shown in Fig.~\ref{fig:p-hcontinuum}. The boundaries of the intraband and the interband portions of the particle-hole excitation continuum are given by $|\Omega | = q$. At $q = 0$, the two boundaries merge at the point $\omega = 2\omega_B + u\Delta n$. We note in passing that this represents a seeming violation of the Larmor theorem, which predicts that the boundaries should merge at $\omega = 2\omega_B$, independent of the interaction. This inconsistency is resolved by noticing that in the full spin-spin polarization function, Fig.~\ref{fig:loop}, the self-energy corrections in the Green's functions should be accompanied with an infinite series of the ladder corrections (the diagrams with no intersection of interaction lines or inner electron loops). A simple geometric series calculation shows that taking the ladder diagram into account amounts to the following modification of the polarization function,
\begin{equation}
\label{Zero momentum1}
{\cal P} (\omega,0)\to \Pi (\omega,0)= \frac{ {\mathcal P}}{1+(u/2)  {\mathcal P}} = \frac{N  \Delta n }{\omega - 2 \omega_{B} },
\end{equation}
which is independent of the interaction $u$ and thus consistent with the Larmor theorem. 

The spin-current-spin-current polarization function, calculated in Appendix~\ref{appendix:current-current}, is given in the  $\Omega < 0$, $|\Omega| > q$, and $\omega_{B} \gg q$ region by
 \begin{align}
 \label{Total current-current correlation}
{\cal K}
(\omega,q) = &\frac{N\omega_{B}^2 \Omega}{2\pi q^2} \bigg(-1+ \frac{|\Omega|}{\sqrt{\Omega^2 - q^2}} \bigg)
									 - 2 N\tilde \Lambda \nonumber\\
						       	 &-i\frac{N\Omega^2}{16\sqrt{\Omega^2 -q^2}} .							 					 
\end{align}
In the above expression $\tilde \Lambda$ is the momentum cut-off for the linear Dirac dispersion, see Appendix~\ref{appendix:current-current} for details. This linear UV divergence of the spin-current polarization function is also present in the charge current correlation function \cite{Sabio2008,Principi2009}. In the expressions for observables, $\tilde\Lambda$ enters only in combination of $u\tilde\Lambda$, as shown in Section~\ref{sec:modes}. For weak interaction, $u\tilde\Lambda\ll 1$, $\tilde\Lambda$ disappers from the results, while for strong interaction, $u\tilde\Lambda\gtrsim 1$, the low-energy model, Eqs.~\eqref{non-interacting hamiltonian}, \eqref{interacting hamiltonian} is not a valid starting point. 

Finally, the expression for the hybrid spin-spin-current polarization function, treated in Appendix~\ref{appendix:spin-current}, is given by 
\begin{equation}
\label{Interband h2}
{\cal H}(\omega,q) = \frac{N\omega_{B}^2}{4\pi q} \bigg(\! -1 + \frac{|\Omega|}{\sqrt{\Omega^2 - q^2}} \bigg)
						       	 -\frac{iNq\Omega}{32\sqrt{\Omega^2 -q^2}}.							 					 
\end{equation}

This polarization function describes the coupling between spin and spin current fluctuations, which must decouple in the $q\to 0$ limit, as dictated by the spin continuity equation. 
Expanding ${\cal H}(\omega,q)$, near $q \approx 0$, we obtain
\begin{equation}
\label{Interband h3}
{\cal H}(\omega,q) = \frac{N\omega_{B}^2 q}{8\pi \Omega^2},
\end{equation}
as expected.

\subsection{Bethe-Salpeter equation}
  \begin{figure}
 \includegraphics[width=3.5in]{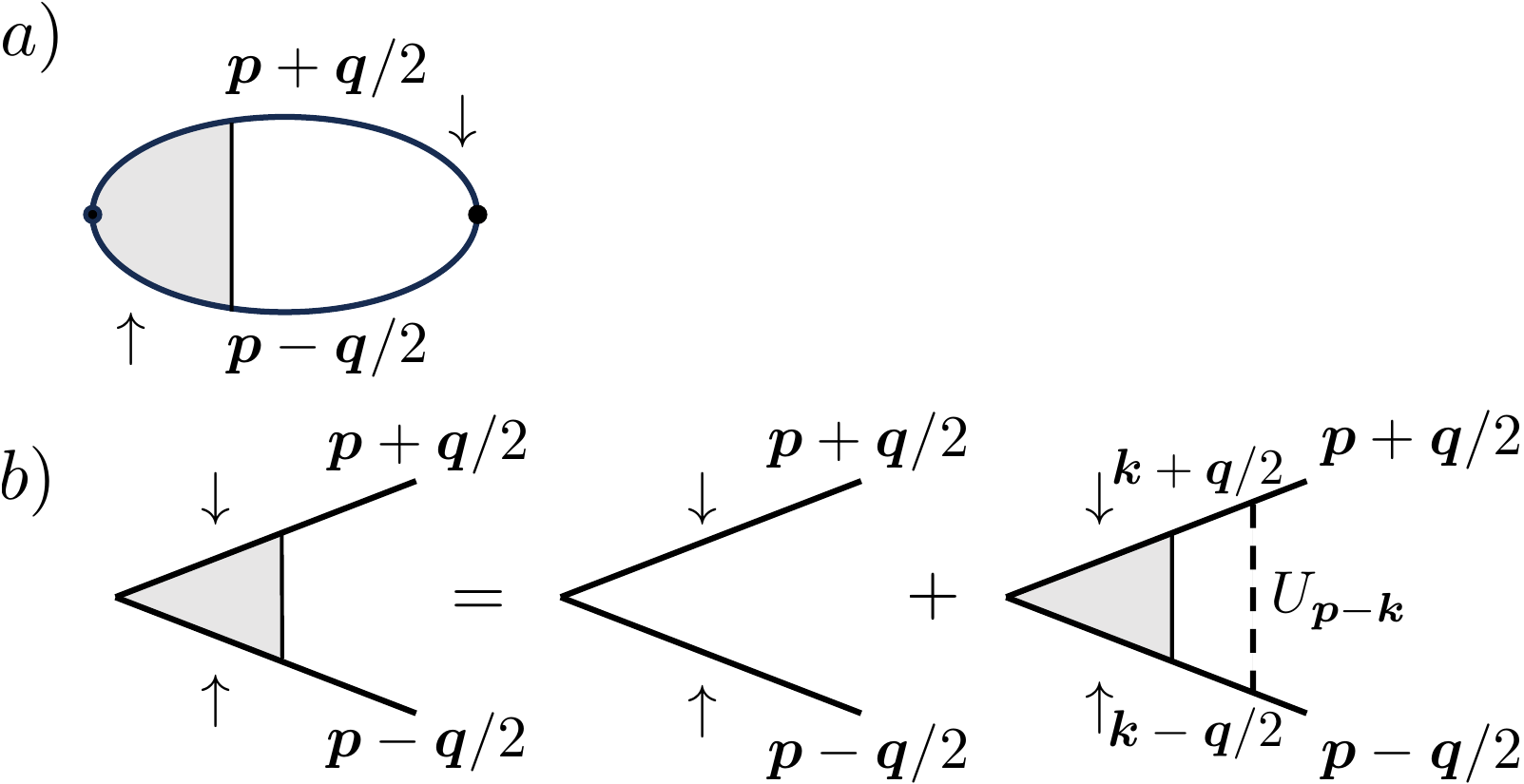}
 \caption{a) Total polarization function, b) vertex summation to all orders in interaction.}
 \label{fig:vertexcorrections}
 \end{figure}
The full spin-spin polarization function, $\Pi$, which determines the transverse spin susceptibility, is obtained by adding vertex corrections to the bubble for $\cal P$, see Figs.~\ref{fig:all-the-loops} and~\ref{fig:loop}. We calculate the vertex corrections in the ladder approximation, which  
amounts to the following modification of the electron loop, illustrated in Figs.~\ref{fig:vertexcorrections}a and~\ref{fig:vertexcorrections}b:
\begin{equation}
\label{Bethe-Salpeter}
 \Pi
 (\omega,{\bf q}) = -i \text{Tr}_\sigma \sum_{{\bf p}}\int \frac{d\epsilon}{2\pi}
 			 \hat {\cal G}^{\downarrow}_{\epsilon_+,\bf{p_+}}\hat \Gamma_{ \omega, {\bf p},{\bf q}} \hat {\cal G}^{\uparrow}_{\epsilon_-,{\bf p_-}}.
 \end{equation}
The vertex function $\hat \Gamma_{ \omega, {\bf p},{\bf q}}$ obeys the Bethe-Salpeter integral equation:
 \begin{equation}
 \label{Gamma1}
 \hat \Gamma_{ \omega, {\bf q}} = 1 +i u\sum_{{\bf k}} \int \frac{d\epsilon}{2\pi}
 							 \hat {\cal G}^{\downarrow}_{\epsilon_+,\bf{k_+}}\hat \Gamma_{ \omega, {\bf q}} \hat {\cal G}^{\uparrow}_{\epsilon_-,{\bf k_-}},
 \end{equation}
graphically shown in Fig.~\ref{fig:vertexcorrections}b.

The independence of the interaction constant $u$ of the momentum $q$ means that the right-hand side in Eq.~(\ref{Gamma1}) does not depend on momentum ${\bf p}$. This ensures that the vertex function  depends only on the transferred frequency $\omega$ and momentum ${\bf q}$. For an isotropic system, we can conclude that the dependence of the vertex function $\hat \Gamma_{ \omega,{\bf q}}$ on the direction of ${ \bf q}$ has the form, 
  \begin{equation}
 \label{Gamma2}
 \hat \Gamma_{ \omega,{\bf q}} = \Gamma_{0}(\omega,q) + \Gamma_{1}(\omega, q)\,{\hat \sigma}_{\bf q}.
 \end{equation}
Substitution of the ansatz (\ref{Gamma2}) into Eq.~(\ref{Gamma1}) leads to two energy-momentum integrals. The first of those integrals is
 \begin{align}
\label{f}	
-2i\sum_{\bf{k}} \int \frac{d\epsilon }{2\pi}\hat {\cal G}^{\downarrow}_{\epsilon_+, {\bf k}_+} \hat {\cal G}^{\uparrow}_{\epsilon_- ,{\bf k}_-}&={\cal P}(\omega,q) +2{\cal H}(\omega,q){\hat \sigma}_{\bf q},
\end{align}
where ${\cal P}(\omega,q)$ is the polarization function obtained in Appendix~\ref{appendix:spin-spin}, and presented in Eq.~\eqref{Non-interacting Polarization}. Function ${\cal H}(\omega,q)$ is found in Appendix C, and Eq.~\eqref{Interband h2}.
 
 The second energy-momentum integral differs from the first on by the presence of the additional spin operator in the integrand:
\begin{align}
\label{g}								
-2i\sum_{\bf{k}}  \int \frac{d\epsilon}{2\pi}\hat {\cal G}^{\downarrow}_{\epsilon_+,\bf{k_+}} {\hat \sigma}_{\bf q} \hat {\cal G}^{\uparrow}_{\epsilon_-, {\bf k_-}}
				& ={\cal K}(\omega,q) {\hat \sigma}_{\bf q}+ 2{\cal H}(\omega,q).
\end{align}
The function ${\cal K}(\omega,q)$ is the spin current correlation function for the non-interacting electrons, see Appendix~\ref{appendix:current-current}, and Eq.~(\ref{Total current-current correlation}) above. Substitution of Eq.~(\ref{Gamma2}) into Eq.~(\ref{Gamma1}) gives the matrix equation (arguments omitted for brevity),
 \begin{align}
2\Gamma_{0} + 2{\hat \sigma}_{\bf q}\Gamma_{1}
 						= 2 - &u ({\cal P} + 2 {\cal H}{\hat \sigma}_{\bf q}) \Gamma_{0}
 						 - u({\cal K}{\hat \sigma}_{\bf q} +2{\cal H})\Gamma_{1},
 \end{align}
which amounts to two coupled scalar equations,
 \begin{eqnarray}
 \Gamma_{0}(2+u {\cal P}) + 2\Gamma_{1}u {\cal H} &=& 2, \nonumber\\
 2\Gamma_{0} u {\cal H} + \Gamma_{1} (2+u{\cal K}) &=& 0,
 \end{eqnarray}
 whose solutions are,
\begin{subequations}
 \begin{align}
 \Gamma_{0} =& \frac{4+4u{\cal K}}{(2+u {\cal P})(2+u{\cal K}) -4u^2 {\cal H}^2}, \\
  \Gamma_{1} =& -\frac{4u {\cal H}}{(2+u {\cal P})(2+u{\cal K}) -4u^2 {\cal H}^2}.
 \end{align}
\end{subequations}

Knowledge of the vertex function leads to the final expression for the polarization function (\ref{Bethe-Salpeter}):
\begin{equation}
 \label{Interacting polarization a}
\Pi
(\omega,q)  = 2 \frac{(2+ u{\cal K}){\cal P} - 4u {\cal H}^2}{(2+u {\cal P})(2+u{\cal K}) -4u^2 {\cal H}^2}.
 \end{equation} 
The polarization function (\ref{Interacting polarization a}) has the same general form as the polarization function obtained in Ref.~\cite{Agarwal2020} for $\omega_B=0$, but the functions ${\cal P}$, $K$, and ${\cal H}$ in Eq.~(\ref{Interacting polarization a}) depend on the magnetic field.
 
In the absence of interactions, $u = 0$, Eq.~(\ref{Interacting polarization a}) reduces to Eq. (\ref{Non-interacting Polarization}). 
 At zero momentum, ${\cal H}(\omega,0) = 0$ (see Eq.~(\ref{Interband h3}) in Appendix C), and ${\cal P}(\omega,0) = N\Delta n/\Omega$, so that the polarization function becomes,
 \begin{equation}
 \label{Zero momentum2}
 \Pi(\omega,0) = \frac{2{\cal P}}{2+u {\cal P}} = \frac{N \Delta n}{\omega - 2\omega_{B}},
 \end{equation}
in agreement with Eq.~(\ref{Zero momentum1}).

\section{Dispersion of collective modes}\label{sec:modes}
The poles of the polarization function in the ladder approximation \eqref{Interacting polarization a} are  determined by the equation,
\begin{equation}
\label{Roots}
(2+u {\cal P})(2+u{\cal K}) -4u^2 {\cal H}^2 = 0,
\end{equation}
yield the spectrum of the collective modes. The functions ${\cal P}$, ${\cal K}$ and ${\cal H}$ are given by Eqs.~(\ref{Non-interacting Polarization}), (\ref{Total current-current correlation}), and (\ref{Interband h2}) respectively.
These expressions for the functions ${\cal P}$, ${\cal K}$, and ${\cal H}$ apply provided that $|\Omega| > q$, and $\omega_{B} \gg q$, which is also the regime where the collective modes exist.

The functions  ${\cal P}$, ${\cal K}$, and ${\cal H}$ satisfy the following relations that follow from Eqs.~(\ref{Non-interacting Polarization}), (\ref{Total current-current correlation}), and (\ref{Interband h2}), 
\begin{eqnarray}
{\cal K}(\omega,q) =\frac{2\Omega}{q} {\cal H}(\omega,q) -  2N\tilde \Lambda,\\
\frac{\Omega}{2q} {\cal P}(\omega,q)-{\cal H}(\omega,q) = \frac{\Delta n}{q}.
\end{eqnarray} 
These relations allows us to simplify Eq.~(\ref{Roots}) to, 
\begin{equation}
\label{Roots_simplified}
2+u [(1-u N\tilde\Lambda) P +{\cal  K} + 2u\Delta n {\cal H}/q] =0,
\end{equation}
where the factor of $1-u N\tilde\Lambda$ originates from the diverging (last) term in \eqref{Current-current Interband b}.  In what follows we assume that the interaction is sufficiently weak, $u N \tilde\Lambda \ll 1$, such that we can replace $1-u N\tilde\Lambda$ with unity. This is a sensible approximation, since the idealized form of the interaction Hamiltonian used in this work, Eq.~\eqref{interacting hamiltonian}, already implies that we can only reveal the structure of the collective mode spectrum, rather than its quantitative details. The limit of weak interaction does not affect the applicability of the Larmor theorem, which holds regardless of the interaction strength, see Eq.~\eqref{Zero momentum2}, \textit{i.e.} it does not affect the frequency of the Silin mode at zero momentum, fixed to be $2\omega_B$. In the opposite limit of very strong interactions, $uN\tilde\Lambda\gtrsim 1$, the low-energy model defined with Eq.~\eqref{non-interacting hamiltonian} and~Eq.~\eqref{interacting hamiltonian} is not a valid starting point of any analysis anyway.

Eq.~(\ref{Roots_simplified}) determines the real part and the imaginary part (attenuation) of the dispersion of the collective modes. We will analyze it for sufficiently weak interaction, $u^2\Delta n\ll1$. In this the attenuation of a collective mode is small compared to its frequency, and one can treat the imaginary part of Eq.~(\ref{Roots_simplified}) as a perturbation. We note that this does \emph{not} imply that the collective  modes are well-defined. We will address this issue below Eq.~\eqref{eq-main}.

\begin{figure}
    \centering
    \includegraphics[width=2.5in]{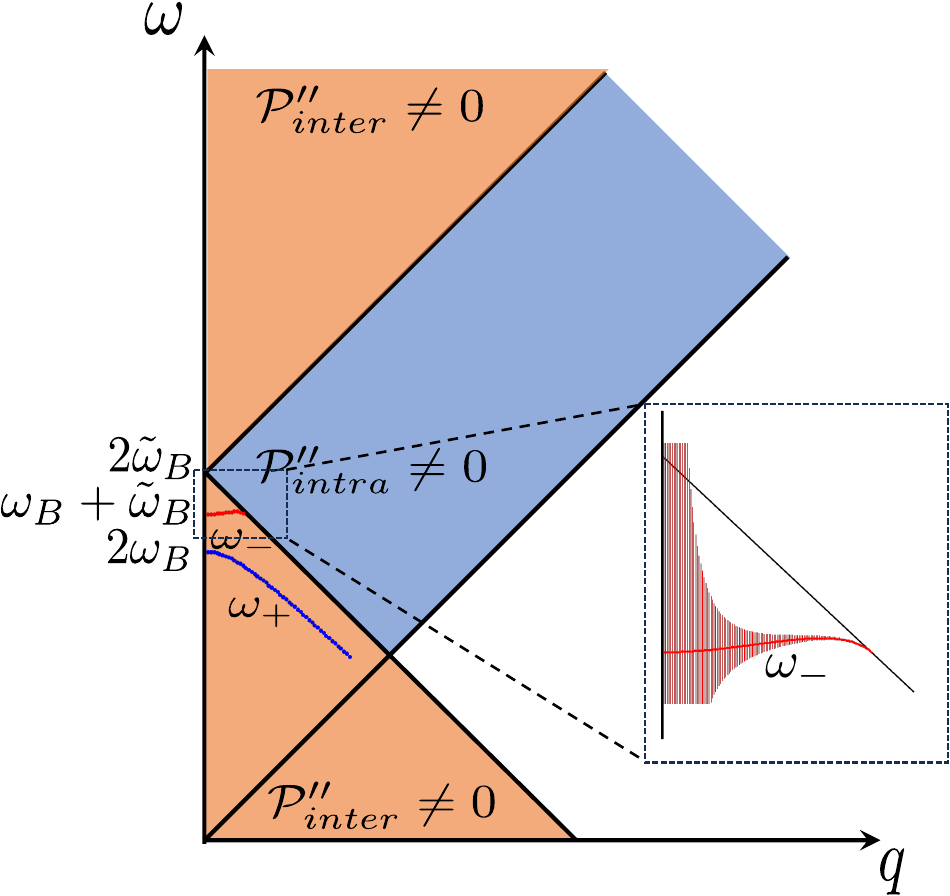}
    \caption{Dispersions of the Silin mode ($\omega_+$, blue line), and the SPS mode ($\omega_-$, red line) for $u\Delta n/2\omega_B=0.25$. Inset: behavior of the SPS mode near the intraband particle-hole continuum boundary. The ``error bars" represent the ratio $\omega_{-}''(q)/(\omega_{-}'(q)-\omega_{-}'(0))$ to arbitrary scale, illustrating that the SPS being overdamped for $q\to 0$, and its progressive sharpening with increasing $q$, see the discussion below Eq.~\eqref{eq-main}.}
    \label{fig:modes}
\end{figure}

Introducing the new dimensionless variables according to (cf.~Appendix E) $z = |\Omega|/u\Delta n $, and $q_0 = q/u\Delta n $, and noting that $\Omega<0$, Eq.~(\ref{Roots_simplified}) reduces to,
\begin{equation}\label{eq:dispersionequation}
\sqrt{z^2- q_{0}^2}(q_{0}^2-1+z) - (q_{0}^2-z+z^2)=0.
\end{equation}
where we have used $N = 2$ for graphene. The corresponding dispersions of the collective modes are shown in Fig.~\ref{fig:modes}.

We can also obtain the behavior of the dispersions at low $q$ analytically. For this purpose it is sufficient to expand Eq.~\eqref{eq:dispersionequation} to the fourth order in $q_0 \ll z$. This yields the following equation,
\begin{equation}
(z-1)(2 z-1) = {q_0^2}\left(1 +\frac{z-1}{4z^2}\right).
\end{equation}
It now follows that two solutions $z_1(q_0)$ and $z_2(q_0)$ exist, which in the limit of $q_0\to 0$, behave as $z_1\to 1$ and $z_2\to 1/2$, respectively. To the second order in $q_0^2$, we find,
\begin{subequations}
\label{collective modes}
 \begin{align}
z_1(q_0) \approx &~1+q_{0}^2, \\
z_2(q_0) \approx &~\frac{1}{2}-\frac{1}{2}q_{0}^2.
 \end{align}
\end{subequations}

The polarization function \eqref{Interacting polarization a} can also be written in terms of the new variables, $z$ and $q_0$, as
\begin{equation}
\label{Interacting polarization b}
\Pi(z,q_{0}) = -\Big(\frac{2}{u}\Big)\frac{\psi(z,q_{0})}{\chi(z,q_0) + i\zeta(z,q_{0})},
\end{equation}
where zeroes of $\chi(z,q_0)  = 0$ corresponds to the roots $z_1$ and $z_2$ in \eqref{collective modes}. 
The function $\psi(z,q_{0})$ determines the residues (spectral weights) of these modes as
\begin{equation}
\phi(z_{i},q_0) =  -\Big(\frac{2}{u}\Big)\frac{\psi(z_i,q_0)(z-z_i)}{\chi(z_i,q_0)},
\end{equation}
where $\psi(z,q_0) = 8z^3( q_0^2+\sqrt{z^2 - q_0^2} - z)/q_0^2$.
For small momentum $q_0 \ll z$, $\psi(z,q_0) \approx 4z^2( 2z -1)$ and
\begin{widetext}
\begin{equation}
\label{widetext pi}
\Pi(z,q_0) \approx -\Big(\frac{2}{u}\Big) \frac{4z^2(2 z-1)}{4z^2(z-1)(2z-1) + q_0^2(1-z-4 z^2) -i\frac{M u }{2}( q_0^2-z+z^2)z^3}. 
\end{equation}
\end{widetext}
The residues of the two poles at $z_1$ and $z_2$ are 
\begin{equation}
\phi(z_{i},q_0)  =  -\Big(\frac{2}{u}\Big) 
					\begin{cases}
					 1-\frac{7}{4}q_0^2, & i = 1 \\
					 q_0^2, & i = 2.
					\end{cases}
\end{equation}
At zero momentum, $q_0=0$, only the $z_1(q_0)$  mode has a nonzero weight, where $z_1\to 1$ (which in the usual variable corresponds to $\omega = 2\omega_B$), with $\Pi(z,0) = -(2/u)/(z-1)$.
This mode is analogous to the Silin spin wave in conventional conductors \cite{Silin1958,Silin1959} and has the downward dispersion, 
\begin{equation}\label{eq:silinmode}
\omega_+ \approx 2\omega_B - \frac{q^2}{u\Delta n},
\end{equation} 
starting at the resonance frequency, $2\omega_B$. The result $\omega_+(0)=2\omega_B$ for the Silin model is general, and does not depend on the assumptions made in this work. 

The second mode $z_2(q_0)$ has the vanishing spectral weight at $q_0\to0$, in agreement with the Larmor theorem. It has the upward dispersion,
\begin{equation}\label{eq:hybridmode}
\omega_- \approx 2\omega_B + \frac{u\Delta n}2 + \frac{q^2}{2u\Delta n }.
\end{equation}
We note that in the present model and for weak interactions, $\omega_-(0)=\omega_B+\tilde\omega_B$, but this result is non-universal. 

For the imaginary parts in the denominator of Eq.~(\ref{Interacting polarization b}), as determined from $\zeta(\omega_{\pm})$, we obtain that,
\begin{equation}
\zeta(z,q_0) = -\frac{u^2\Delta n }{2}( q_0^2-z+z^2)z^3.
\end{equation}
The polarization function for small momenta $q \ll \omega_B$ is thus,
\begin{align}
&\Pi_{}(\omega,q) = \frac{2 \Delta n}{\omega - \omega_+ + i\frac{u  q^2}{4}}  \nonumber\\
&+ \frac{2q^2}{u^2 \Delta n [\omega - \omega_- + i\frac{u^3 \Delta n^2 }{64 }]}.
\label{eq-main}
\end{align} 

We finish this Section with a discussion of attenuation of the two modes. A mode is well-defined if one can build a wave packet that propagates by at least a wavelength before the mode decays. For a general dispersion $\omega(q)=\omega'(q)-i\omega''(q)$, a well-defined mode satisfies 
\begin{align}\label{eq:definedmode}
    \left|\frac{d\omega'(q)}{dq}\right|\cdot\frac{1}{\omega''(q)}>\frac{1}{q}.
\end{align}
For the Silin-type mode with $\omega_+(q)=2\omega_B-q^2/u\Delta n-iuq^2/4$, condition~\eqref{eq:definedmode} implies that the mode is well defined for $u\Delta n^{1/2}<1$, which defines the weak-interaction limit in which all the equations above were obtained.

In turn, the dispersion of the spin-current mode has a Landau damping with a finite rate at $q\to 0$: $\omega_-=2\omega_B+u\Delta n/2 +q^2/2u\Delta n-i u^3\Delta n^2/64$, hence this mode is overdamped at small $q$. The condition~\eqref{eq:definedmode} determines the value of $q=Q$ for which the mode becomes well-defined: $Q\sim u^2\Delta n^{3/2}$. The spectrum of the spin-current mode ends at $q=u\Delta n/2$. This can be seen from Eq.~\eqref{eq:dispersionequation} by substituting $z=q_0$ in it, which yields the point where the dispersion touches the boundary of the intraband continuum of spin-flip particle-hole excitations. This value of $q$ is parametrically larger than $Q$ for weak interaction, $u\Delta n^{1/2}<1$.

\section{Summary and Conclusions}\label{sec:discussion}

Eq.\eqref{eq-main} summarizes our main finding: transverse spin susceptibility of the magnetized graphene supports two collective excitations that live below the intra-band particle-hole continuum. The main of these collective modes - the Larmor-Silin mode - contributes most to $\Pi_{}(\omega,q) $ and is the well known feature of every magnetized conductor, single band or not. Its existence is guaranteed by the spin-rotational symmetry of the electron Hamiltonian and the Larmor theorem that follows from that. The second, spin-current mode (with the dispersion $\omega_{-}(q)$) is the key new finding of our work. Its existence is not symmetry-protected and is therefore particular to graphene. 

While both of the collective spin modes disperse quadratically, the spin-current mode is less well-defined. According to \eqref{eq-main}, its imaginary part at $q\to 0$ is finite and controlled by interaction, $\propto u^3 \Delta n^2$. Therefore at small $q$ this mode will appear as a non-dispersing resonance. Nonetheless, its intensity increases with $q$ and hence the spin-current contribution to $\Pi_{}(\omega,q) $ increases with $q$ too, which makes it a measurable feature of the dynamical response function.

It would be very interesting to look for the spin-current mode experimentally with the help of optical THz technique or something similar. Up to date, experimental attempts to probe spin sector of graphene are essentially nonexistent. We are aware of the magnetotransport study \cite{Chiappini2016} where the in-plane magnetic field up to 30 T has been applied. It appears that the field-induced Zeeman splitting of electron bands, which the authors of the study estimate to be of the order 3 meV, remains well masked by electrostatic potential fluctuations. 
We hope our work may provide additional motivation for more determined experimental efforts in this promising research direction.

Our findings bear surprising similarity with the dynamical spin response of the magnetized Heisenberg spin-1/2 chain studied in \cite{Keselman2020,Wang2022,Povarov2022}. The similarity concerns the existence of the two collective modes - Larmor and spin-current - in the small momentum region of the transverse spin susceptibility. In the one-dimensional setting the properties of these two modes are analytically determined by the Kac-Moody algebra of the emergent low-energy generators of spin rotations \cite{Wang2022}. Their location on the energy axis (at $q=0$) and their dispersions were also confirmed numerically, and were later measured in the electron spin resonance experiment \cite{Povarov2022} that succeeded in extracting the strength of the interaction between spin fluctuations of the Heisenberg chain.

In this regard, we would like to point out one more promising connection to the physics of two-dimensional frustrated spin models. It has to do with the Dirac quantum spin liquid which has been argued theoretically to realize in quantum Heisenberg model on frustrated kagome and triangular lattices \cite{Ran2009}. Recent theoretical studies \cite{Sherman2023} and intriguing experiment \cite{Xu2023} has renewed interest in physical properties of this exotic spin liquid state. Our investigation of the dynamical spin susceptibility of the magnetized graphene corresponds, in the language of \cite{Ran2009}, to that of the Fermi-pocket state of the magnetized Dirac spin liquid. 

\section{Acknowledgements}
O.A.S. thanks L. Balents, L. Glazman, and D. Maslov for numerous discussions, insightful remarks, and probing questions, and also                 E. Henriksen for pointing out experimental study \cite{Chiappini2016}. D.A.P. is grateful to M. Katsnelson for useful discussions. O.A.S. is supported by the NSF CMMT program under Grant No. DMR-1928919. The work of D.A.P. was supported by the National Science Foundation Grant No. DMR-2138008.

\appendix

\section{Spin-spin polarization function}\label{appendix:spin-spin}

In this and subsequent Appendix Sections, we derive all the quantitative results of this work for a single Dirac point. In the main text, however, we will recover $N = 2$ accounting for the two Dirac points in graphene. Since we are proposing a new collective mode in a well-known system, we show the pertinent calculations in great detail. 

For brevity, the calculations are performed for non-interacting electrons. The polarization functions of interacting electrons are then obtained  by including the electron self-energy, which simply amounts to the renormalization of the Zeeman energy, $\omega_B\to \tilde\omega_B$, as described above in relation to Eq.~(\ref{eq:tildeomega}). 

We start with the transverse spin polarization function of a single Dirac point, given by 
\begin{equation}\label{Polarization a}
 {\mathcal P}(\omega,\bm{q}) = -i \text{Tr}_\sigma \sum_{\bm{k}} \int\frac{d\epsilon}{2\pi}  \hat {\cal G}^{\downarrow}_{\epsilon_{+},\, {\bf k}_+} \hat {\cal G}^{\uparrow}_{\epsilon_{-}, \,{\bf k}_-},
\end{equation}
where $\epsilon_{\pm} =\epsilon \pm \omega/2$,  ${\bf k}_\pm ={\bf k}\pm{\bf q}/2$, and the Green's function of graphene is written in pseudo-spin space as,
\begin{equation}
\label{Green's function}
\hat {\cal G}^{\tau}_{\epsilon,\bm{k}} = \frac{1}{2} \sum_{s = \pm1} \frac{1+s \hat \sigma_{\bm{k}}}{\epsilon- {\cal E}^{\tau}_{s}(k) +is0^+}.
\end{equation}
In this expression, $\hat \sigma_{{\bf k}} = \hat {\bm \sigma} \cdot {\bf k}/k$ is the projection of the pseudospin Pauli matrix onto the direction of the electron momentum ${\bf k}$ and $\tau$ can take one of the two values, $\uparrow$ or $\downarrow$. 
Integration over the energy variable $\epsilon$ yields
\begin{equation}
 \label{Polarization b}
\! \! \!\ {\mathcal P}
(\omega,\bm{q})\! = \sum_{s,s'}\! \sum_{\bf{k}} \! \frac{(n_{s'}^{\uparrow}(k) - n_{s}^{\downarrow}(|{\bf k}+{\bf q}|)F_{ss'}(\bm{k},{\bf q})}{\omega- {\cal E}_{s}^{\downarrow}(|\bm{k}+{\bm q}|)+{\cal E}_{s'}^{\uparrow}({k}) + i0^+},
\end{equation}
where the factor $F_{ss'}({\bf k, q})=|\langle {\bf k}+{\bf q}, s|{\bf k}, s' \rangle|^2 $ characterizes the overlap of the band states,
\begin{align}
\label{Form factor Fa}
F_{ss'}({\bf k, q}) &= \frac{1}{4}\text {Tr}[(1+s' \hat \sigma_{\bm{k}}) (1 + s\hat \sigma_{\bm{k+q}})] \nonumber\\
				&=  \frac{1}{2}(1+s s' \cos\theta_{\bm{k}, \bm{k+q}}).
\end{align}
Here $\theta_{{\bf k,k+q}}$ is the angle between vectors ${\bf k}$ and ${\bf k}+{\bf q}$ and $n^\tau_s(k) = (\exp{({\cal E}^\tau_s(k)/T)} + 1)^{-1}$ is the Fermi-Dirac distribution.
The  factor $F_{ss'}({\bf k, q})$ accounts for the chiral properties of graphene and is responsible for the suppression of backscattering processes. For intrinsic graphene with no magnetic field, $F_{ss'}({\bf k, q})$ vanishes for $\theta_{\bm{k}, \bm{k+q}} = \pi$ for intraband transitions, $s = s'$, and also vanishes for $\theta_{\bm{k}, \bm{k+q}} = 0$ for interband transitions, $s = -s'$. The trace of linear in $\hat \sigma_{\bm{k}}$ terms vanishes.  

To calculate the integral in Eq. (\ref{Polarization b}), we  choose the absolute values $k$ and $p =|{\bf k}+{\bf q}|$ as the new variables, to replace the angle variable $\theta_{\bf k}$ in the integral over $d^2k =k \,dk \,d\theta_{\bf k}$ with  $p$. To make use of the new variables, we first notice the identity,
\begin{equation}
\label{Identity1}
\int\limits_0^\infty dp~\frac{p}{kq} \delta\Big(\frac{p^2 - k^2 - q^2}{2kq} -\cos \theta_{\bf k} \Big) = 1. 
\end{equation}
The  integral over $d \theta_{{\bf k}}$  now leads to the following expression,
\begin{align}
\label{Identity2}
\int\limits_{0}^{2 \pi}  d\theta_{\bf k} ~& \delta\Big(\frac{p^2 - k^2 - q^2}{2kq} -\cos \theta_{\bf k} \Big) \nonumber\\
			&= \frac{4kq}{\sqrt{((p+k)^2 - q^2)(q^2-(p-k)^2)}},
\end{align}
provided that $|p-k|<q<p+k$ (and zero otherwise).
The polarization function, Eq. (\ref{Polarization b}), now reduces to, 
 \begin{align}
 \label{Polarization c}
 {\mathcal P}
 (\omega,\bm{q}) &= 4 \int  \frac{dk dp }{(2\pi)^2} \frac{k p}{\sqrt{((k+p)^2 - q^2)(q^2-(k-p)^2)}}  \nonumber\\
							& \times\frac{(n_{s'}^{\uparrow}(k) - n_{s}^{\downarrow}(p))}{\omega- {\cal E}_{s}^{\downarrow}(p)+{\cal E}_{s'}^{\uparrow}({k}) + i0^+}F_{ss'}({\bf k, q}).
\end{align}
Using the argument of the delta-function in Eq.~(\ref{Identity1}) to express $\cos \theta_{\bf k}$, we obtain the factor $F_{s,s'}$ via $p$ and $k$,
\begin{eqnarray}
\label{Form factor Fb}
F_{s,s}({\bf k, q}) &=& \frac{(k+p)^2 - q^2}{2 k p},\nonumber\\
F_{s,-s}({\bf k, q}) &=& \frac{q^2 - (p-k)^2 }{2 k p}.
\end{eqnarray}
Both the intraband transitions, $s =s'$, and the interband transitions, $s' =-s$, contribute to the polarization function (\ref{Polarization c}). Below we will look at these contributions to both the real and imaginary part of the polarization function.

\subsection{Intraband transitions}
Let us first consider the intraband terms, $s =s'$, in Eq.~(\ref{Polarization c}).  Two processes where an electron with momentum $\bf{k}$ and spin $\uparrow$ absorbs external momentum $\bf{q}$ and transitions to the state with momentum $\bf{k+q} = \bf{p}$ and spin $\downarrow$ exist. The corresponding energy change is ${\cal E}_{s}^{\uparrow}(\bm{k}) \rightarrow {\cal E}_{s}^{\downarrow}(\bm{p})$. The two contributions correspond to the transitions within the lower cone $s = -1$, where ${\cal E}_{-1}^{\downarrow}(\bm{p})  -  {\cal E}_{-1}^{\uparrow}(\bm{k})  =  2\omega_{B} - vp + vk $,  and within the upper cone $s = 1$, where ${\cal E}_{1}^{\downarrow}(\bm{p})  -  {\cal E}_{1}^{\uparrow}(\bm{k})  =  2\omega_{B} + vp - vk $. Note that $n_{-1}^\uparrow (k) - n_{-1}^\downarrow (p)= \Theta(\omega_{B} - p) $. The presence of the Heaviside step function $\Theta(\omega_{B} - p)$ indicates that the momentum of the final state must be less than $\omega_B$. Similarly, for the upper cone we have, $n_{1}^\uparrow (k) - n_{1}^\downarrow (p) =  \Theta(\omega_{B} - k)$,  indicating that the initial state must have momentum less than $\omega_B/\hbar v$. The contribution of the intraband transitions into the polarization function (\ref{Polarization c}) then reduces to,
\begin{align}
\label{Intraband polarization a}
&{\mathcal P}_{intra}
(\omega,q) = \int\limits_{0}^{\infty}  \frac{dk ~dp}{(2\pi)^2}  \sqrt{\frac{(k+p)^2 - q^2}{q^2-(k-p)^2}} \nonumber\\
									& \times \bigg(\frac{\Theta(\omega_{B} - p)}{\Omega  + p - k + i0^+}
							 + \frac{\Theta(\omega_{B} - k)}{\Omega - p + k  + i0^+} \bigg),
\end{align}
where the notation $\Omega = \omega - 2\omega_B$ is used for the shifted frequency, measured from $2\omega_B$. Because in the remaining integrals over $dp\,dk$ in Eq.~(\ref{Intraband polarization a}), the integrand depends only on the sum $k+p$ and the difference $k-p$, rotation of the integration variables, $k+p=qx $, and $ k-p=qy$, allows one to factorize the integrals as follows,
\begin{equation}
\label{Intraband polarization b}
\!{\mathcal P}_{intra}
(\omega,q) = q^2 \! \int\limits_{0}^{\infty}  \frac{dx dy}{(2\pi)^2}  \sqrt{\frac{x^2 - 1}{1-y^2}} \frac{\theta(2\omega_{B} - q(x-y))}{\Omega - qy + i0^+}.
\end{equation}
The imaginary part of the polarization function arises from the $\delta (\Omega -qy)$ contribution to the integral in Eq.~(\ref{Intraband polarization b}),
\begin{align}
\label{Intraband polarization im a}
{\mathcal P}_{intra}
''(\omega,q) = &-q^2 \! \int\limits_{1}^{\infty}  \frac{dx }{4\pi} \int\limits_{-1}^{1 } dy  \sqrt{\frac{x^2 - 1}{1-y^2}} \theta(2\omega_{B} - q(x-y)) \nonumber\\
			&\times \delta(\Omega - qy).
\end{align}
Assuming that both $\omega>0 $ and $ q>0$, the $y-$integral eliminates the delta-funciton and is non-zero only for $q>|\Omega|$. The remaining $x-$integral is calculated in Appendix D and is non-zero for $\omega > q$,
\begin{align}
\label{Intraband polarization im b}
 {\mathcal P}_{intra}
 ''(\omega,q) = -&\frac{1}{8\pi \sqrt{q^2-\Omega^2}} \Big(\omega \sqrt{\omega^2 - q^2 } \nonumber\\
											& -q^2\ln\big(\frac{\omega + \sqrt{\omega^2 -q^2}}{q}\big)\Big).
\end{align}

Real electron-hole excitations exist in the range of the $\omega,q$ space where the polarization function is non-zero (indicated by shading in Fig.~\ref{fig:p-hcontinuum}). This range is defined by the following conditions: $\omega>q> |\Omega|$. In the absence of magnetic field, $\omega_{B} = 0$, $\Omega = \omega$, the imaginary part of the polarization function vanishes since no intraband transitions are possible in intrinsic graphene at zero magnetic field and zero temperature.

The real part of the polarization function is given by the principal value of the integral in Eq. (\ref{Intraband polarization b}),
\begin{equation}
 \label{Intraband polarization re a}
 {\mathcal P}_{intra}
 '(\omega,q) = q^2 P\int\limits_{1}^{\infty}  \frac{dx \sqrt{x^2 - 1}}{(2\pi)^2}  \int\limits_{-1}^{1} dy \frac{\Theta(2\omega_{B} - q(x-y))}{\sqrt{1-y^2}(\Omega - qy)} .
\end{equation}
For  values of the external momentum that are small compared with the magnetic field, $q \ll \omega_B$, the term $qy$ in the argument of the step function can be neglected and the two integrals in Eq.~(\ref{Intraband polarization re a}) factorize. Large arguments are important in the $x$-integral, which simply yields a factor of $2\omega_B^2/q^2$. The remaining integral
\begin{equation}
 \label{Intraband polarization re b}
 {\mathcal P}_{intra}
 '(\omega,q) = \frac{\omega_B^2}{2\pi^2} \int\limits_{-1}^{1} dy\frac{1}{ \sqrt{1-y^2}(|\Omega| + qy)}
\end{equation}
 can be evaluated by the substitution, $y = \cos\phi$, and is non-zero for $|\Omega| >q$ (see Appendix D for  details). The contribution of the intraband transitions to the real part of the polarization function is thus
\begin{equation}
 \label{Intraband polarization re c}
 {\mathcal P}_{intra}
 '(\omega,q) = -\frac{\omega_{B}^2}{2\pi\sqrt{\Omega^2 - q^2}}.
\end{equation}
It is proportional to the second power of the magnetic field, $\omega_{B}^2$, which is the measure of the number of electrons and holes produced by the magnetic fiel. 

\subsection{Interband transitions}
The interband transitions, $s = -s'$, in Eq.~(\ref{Polarization c}) correspond to the processes where a spin-up electron with momentum $\bf{k}$ and energy ${\cal E}_{s}^{\uparrow}({k})$ absorbs external momentum $\bf{q}$ and transitions to the spin-down state with momentum $\bf{k+q} = \bf{p}$ and energy ${\cal E}_{-s}^{\downarrow}({p})$. For the transitions from the lower cone $s=-1$ to the upper cone $s=1$, the energy change is  ${\cal E}_{1}^{\downarrow}({p})  -  {\cal E}_{-1}^{\uparrow}({k})  =  2\omega_{B} + vp + vk $ and $n_{-1}^{\uparrow}(k) - n_1^{\downarrow} (p) = 1 $. For the transition from the upper cone $s = 1$ to the lower cone $s = -1$, the energy change is  ${\cal E}_{-1}^{\downarrow}({p})  -  {\cal E}_{1}^{\uparrow}({k})  =   2\omega_{B} - vp - vk$ and $n_{1}^{\uparrow}(k) - n_{-1}^{\downarrow} (p) = \Theta(\omega_{B} - k) - \Theta(p - \omega_{B}) $. The latter transitions are possible if the initial state has momentum less than $\omega_B$ and the final state has momentum greater than $\omega_B$. The contribution from the interband transitions to the polarization function  Eq.~(\ref{Polarization c}) is
\begin{align}
\label{Interband polarization a}
& {\mathcal P}_{inter}
(\omega,q) = \int  \frac{dk dp }{(2\pi)^2} \sqrt{\frac{ q^2 - (p-k)^2 }{(k+p)^2-q^2)}} \nonumber\\
									& \times \bigg(\frac{1}{\Omega - p - k + i0^+}
									  + \frac{\Theta(\omega_{B} - k) - \Theta(p - \omega_{B})}{\Omega+ p + k  + i0^+} \bigg).
\end{align}
Introducing the rotated variables, $k+p=qx $, and $ k-p=qy$, the integrals are factorized as follows,
\begin{align}
\label{Interband polarization b}
 {\mathcal P}_{inter}
 &(\omega,q) = \frac{q^2}{2}  \int \frac{dx dy}{(2\pi)^2}  \sqrt{\frac{1-y^2}{x^2 - 1}}
									 \bigg(\frac{1}{\Omega - qx + i0^+} \nonumber\\
									&+\frac{\Theta(2\omega_{B} - q(x-y)) - \Theta(q(x+y) - 2\omega_{B})}{\Omega + qx + i0^+} \bigg).
\end{align}

The real and imaginary parts of this expression can be obtained in a way that is similar to the one used above for the case of the intraband transitions, cf.~Eq.~(\ref{Intraband polarization b}). In particular, the imaginary part of the polarization function arises from the delta-functions associated with the poles at $\Omega =qx$ and is given by,
\begin{align}
\label{Interband polarization im a}
{\mathcal P}_{inter}
'' &(\omega,q) = -\frac{q^2}{8\pi \sqrt{\Omega^2 - q^2}}  \int\limits_{-1}^{1}dy \sqrt{1-y^2}
											\bigg(\Theta(\Omega)\Theta(\Omega -q) \nonumber\\
											&+ \Theta(-\Omega)\theta(|\Omega| -q)(\Theta(qy + \omega) - \Theta(qy - \omega)) \bigg).
\end{align}
The remaining integral is elementary,
\begin{align}
\label{Interband polarization im b}
&{\mathcal P}_{inter}
''(\omega,q) =-\frac{1}{8\pi \sqrt{\Omega^2 - q^2}} \Bigg[\Theta(\Omega)\Theta(\Omega -q)\frac{\pi q^2}{2}   \nonumber\\
											& +\Theta(-\Omega)\Theta(|\Omega| -q)\Bigg(\Theta(\omega -q) \frac{\pi q^2}{2}  \nonumber\\
											& + \Theta(q - \omega)\bigg( \omega \sqrt{q^2-\omega^2} + q^2\arcsin(\frac{\omega}{q}) \bigg)\Bigg) \Bigg].
\end{align}

The imaginary part ${\mathcal P}_{inter}''$ is non-zero in two regions where the electron-hole excitations can exist: a) $\Omega>0, \Omega>q$ and b) $\Omega <0, \Omega<-q$, indicated by the orange shading in Fig.~\ref{fig:p-hcontinuum}. In the latter region (where, as shown below, collective excitations reside), the imaginary part is (also assuming $q<\omega$),
\begin{equation}
{\mathcal P}_{inter}
''(\omega,q) = -\frac{q^2}{16\sqrt{\Omega^2 - q^2}} .
\end{equation}

The contribution of the interband transitions to the real part of the polarization function is given by the principal value of the integral in Eq.~(\ref{Interband polarization b}).
For $\omega_{B} \gg q$, we can neglect $y$ in the  step functions in the second term since $y \in (-1,1)$. The integrals over $x$ and $y$ factorize, with the $y$ integral simply bringing the factor $\pi/2$,
\begin{align}
\label{Interband polarization re a}
 {\mathcal P}_{inter}
 '(\omega,q) = \frac{q^2}{8\pi}  
									 \bigg({\text P}\int\limits_{1}^{\infty} dx\frac{qx }{\sqrt{x^2 -1}(\Omega^2 - q^2 x^2)} \nonumber\\
										+ {\text P}\int\limits_{1}^{{2\omega_{B}}/{q} } \frac{dx}{\sqrt{x^2 -1}(\Omega+qx )}  \bigg).
\end{align}
The first integral is non-zero provided that $q > |\Omega|$. Assuming that $\omega_B \gg q$, the upper limit in the second integral can also be extended to $\infty$. The second integral is calculated in the Appendix D. In the region of interest for the collective modes, $-q<\Omega < 0$, the real part becomes,
\begin{align}
\label{Interband polarization re b}
 {\mathcal P}_{inter}
 '(\omega,q) = -\frac{q^2}{16\pi\sqrt{\Omega^2 -q^2}}  \ln\frac{2|\Omega|}{q} .
\end{align}
Note that the interband contribution into the polarization function is small compared with the intraband contribution, ${\mathcal P}_{inter}/{\mathcal P}_{intra} \sim q^2/\omega_B^2$ and may be ignored in the  calculations of the dispersion relation of collective excitations, performed in section IV.

The above calculations have been performed for a single Dirac point. The obtained polarization function, including both the real and imaginary parts and the contributions from intraband and interband transitions, $ {\mathcal P}= {\mathcal P}_{intra}+ {\mathcal P}_{inter}$, has the following form in the region $\Omega < 0$, $|\Omega| > q$, and $\omega_{B} \gg q$,
\begin{equation}
\label{Non-interacting Polarization-appendix}
 {\mathcal P}
 (\omega,q) = -\frac{N \omega_{B}^2}{2\pi\sqrt{\Omega^2 - q^2}}
										-i\frac{N q^2}{16\sqrt{\Omega^2 - q^2}},
\end{equation}
where we introduced the number $N$ of the Dirac points ($N =2$, for graphene).

\section{Spin-current-spin-current polarization function}\label{appendix:current-current}

The spin current correlation function is given by the expression,
\begin{equation}
\label{Current-current correlation a}
{\cal K}
(\omega,\bm{q}) = -i \text{Tr}_\sigma \sum_{\bm{k}} \int\frac{d\epsilon}{2\pi}\hat {\cal G}^{\downarrow}_{\epsilon_{+}, {\bf k}_+}\sigma_x \hat {\cal G}^{\uparrow}_{\epsilon_{-} ,{\bf k}_-}\sigma_x.
 \end{equation}
 Integrating over the energy variable, we obtain,
\begin{equation}
\label{Current-current correlation b}
{\cal K}
(\omega,\bm{q}) = \!\sum_{s,s'}\! \sum_{\bf{k}} \! \frac{(n_{s'}^{\uparrow}(k) - n_{s}^{\downarrow}(|{\bf k}+{\bf q}|)G_{ss'}(\bm{k},{\bf q})}{\omega- {\cal E}_{s}^{\downarrow}(|\bm{k}+{\bm q}|)+{\cal E}_{s'}^{\uparrow}({k}) + i0^+}.
\end{equation}
Note that the form of this expression is similar to that of the polarization function in Eq.~(\ref{Polarization b}) with the factor $F_{ss'}({\bf k, q})$ replaced by $G_{ss'}({\bf k, q})$,
\begin{align}
\label{Form factor Ga}
G_{ss'}({\bf k, q}) &= \frac{1}{4}\text {Tr}[(1+s' \hat \sigma_{\bm{k}}) \sigma^x (1 + s\hat \sigma_{\bm{k+q}})\sigma^x] \nonumber\\
				&=  \frac{1}{4}(2+s s' \text{Tr}(\sigma_i \sigma_x \sigma_j \sigma_x) \hat k_{i}(\hat k + \hat q)_{j}).
\end{align}
The trace in the second term gives, $\text{Tr}(\sigma_i \sigma_x \sigma_j \sigma_x)\hat k_{i}(\hat k + \hat q)_{j} = 2[\hat k_x(\hat k + \hat q)_x - \hat k_y(\hat k + \hat q)_y] = 2(\cos \theta \cos \beta - \sin \theta \sin \beta)$, where $\theta$ and $\beta$ are, respectively, the angles ${\bf k}$ and ${\bf k+q}$ make with ${\bm q}$. 
\begin{align}
\label{Form factor Gb}
G_{ss'}({\bf k, q}) &=  \frac{1}{2}\bigg(1+s s' \frac{k(2\cos^2{\theta}-1)+q\cos{\theta}}{|\bm{k+q}|}\bigg).
\end{align}
The integral can be calculated using the same method as above, by taking absolute values $k$ and $p =|{\bf k}+{\bf q}|$ as the new variables and replacing the angle variable $\theta_{\bf k}$ in the integral over $d^2k =k \,dk \,d\theta_{\bf k}$ with  $p$.

The factor $G_{s,s'}$ can be written in terms of the factor $F_{s,s'}$ (see Eq.~(\ref{Form factor Fa})) as,
\begin{eqnarray}
\label{Form factor Gc}
G_{s,s'}({\bf k, q}) &=& \frac{(p-ss' k)^2}{q^2} F_{s,s'}({\bf k, q}).
\end{eqnarray}

The contribution of the intraband transitions, $s' = s$, into the spin current correlation function is
\begin{align}
\label{Current-current Intraband a}
{\cal K}_{intra}
(\omega,q) & = \frac{q^2}{4\pi^2}  \int\limits_{1}^{\infty}dx \sqrt{x^2-1} 
 				\int\limits_{-1}^{1} dy \frac{y^2}{\sqrt{1-y^2}} \nonumber\\
 				& \times \frac{\theta(2\omega_{B} - q(x-y))}{\Omega - qy + i0^+} .			
\end{align}
In the region where collective modes exist, $\Omega < 0$ and $|\Omega| > q$, the imaginary part of ${\cal K}(\omega,q)$ is zero. For the real part, we get,
\begin{align}
\label{Current-current Intraband b}
{\cal K}_{intra}
(\omega,q)  = &\!\int\limits_{1}^{2\omega_{B}} \! \frac{dx}{4\pi^2}\sqrt{x^2 -q^2} 
 				\int\limits_{-1}^{1}\! \frac{ dy~y^2}{\sqrt{1-y^2}(\Omega +qy)}. 	
 \end{align}
 The $x-$integral brings the factor $2\omega_B^2/q^2$. The $y$-integral is calculated in the Appendix D and gives,
 \begin{equation}
 \label{Current-current Intraband c}
 {\cal K}_{intra}
 (\omega,q)  = \frac{|\Omega|}{8\pi}\Big(\frac{2\omega_{B}}{q}\Big)^2 
 					\bigg(1 - \frac{|\Omega|}{\sqrt{\Omega^2 -q^2}} \bigg).
 \end{equation}

For the interband transitions, $s' = -s$, we can write,
\begin{align}
\label{Current-current Interband a}
{\cal K}_{inter}
(\omega,q)  = \frac{1}{8\pi^2q^2}  \int\limits_{q}^{\infty}dx \frac{x^2}{\sqrt{x^2-q^2}} 
 				\int\limits_{-q}^{q}dy \sqrt{q^2-y^2} \nonumber\\
 			\Big(\frac{1}{\Omega - x}
			+ \frac{\theta(2\omega_{B}-x+y) - \theta(x+y-2\omega_{B})}{\Omega+x}\Big).			
\end{align}
The above integral is power-law divergent and is cut-off at the upper limit by the momentum cutoff $\Lambda$. This divergence can be most easily evaluated by separating the $\Omega = 0$ contribution: 
\begin{align}
\label{Current-current Interband b}
&{\cal K}_{inter}
(\omega,q)  = -\frac{|\Omega|}{8\pi^2q^2} \int\limits_{q}^{\infty}\! dx \frac{x}{\sqrt{x^2-q^2}} \int\limits_{-q}^{q}dy \sqrt{q^2-y^2} \nonumber\\
			&\Big(\frac{1}{\Omega - x} - \!\frac{\theta(2\omega_{B}-x+y) - \theta(x+y-2\omega_{B})}{\Omega+x}\Big)
			\! - 2N \tilde \Lambda,
\end{align}
where, $\tilde \Lambda =(\Lambda - \omega_{B})/8\pi$. The linear UV divergence of the spin current correlation function, \eqref{Current-current Interband b}, results from the linear relativistic dispersion. It is also present in the charge current correlation function \cite{Sabio2008,Principi2009}. 

The first integral above is regular and has the form similar to Eq. (\ref{Interband h1}).
The total current-current correlation function, $ {\mathcal K}= {\mathcal K}_{intra}+ {\mathcal K}_{inter}$, that combines both the intraband \eqref{Current-current Intraband c} and the interband \eqref{Current-current Interband b} contributions in region $\Omega < 0$, $|\Omega| > q$, and $\omega_{B} \gg q$ is
 \begin{align}
 \label{Total current-current correlation-appendix}
{\cal K}
(\omega,q) = &\frac{N\omega_{B}^2 \Omega}{2\pi q^2} \bigg(-1+ \frac{|\Omega|}{\sqrt{\Omega^2 - q^2}} \bigg)
									 - 2 N\tilde \Lambda \nonumber\\
						       	 &-i\frac{N\Omega^2}{16\sqrt{\Omega^2 -q^2}} .								 \end{align}
where we introduced the number $N$ of the Dirac points ($N =2$, for graphene).
\section{Spin-spin-current polarization function}\label{appendix:spin-current}
The hybrid spin-spin-current polarization function is given by
\begin{equation}
{\cal H}(\omega,\bm{q}) = - \frac{i}{2}  \text{Tr}_\sigma \sum_{\bm{k}} \int\frac{d\epsilon}{2\pi} \hat G^{\uparrow}_{\epsilon_{-} ,{\bf k}_-}\sigma^x \hat G^{\downarrow}_{\epsilon_{+}, {\bf k}_+}.
 \end{equation}
Integration over the energy variable gives,
\begin{equation}
{\cal H}(\omega,\bm{q}) =  \sum_{s,s'} \int \frac{d\bm{k} }{(2\pi)^2} \frac{(n_{s'}^{\uparrow}({\cal E}_{\bm{k_-}}) - n_{s}^{\downarrow}({\cal E}_{\bm{k_+}}))h_{ss'}({\bf k, q})}{\omega- {\cal E}^{s}_{\downarrow}(\bm{k_+})+{\cal E}^{s'}_{\uparrow}(\bm{k_-}) + i0^+}.
\end{equation}
We note that the form of the above equation is identical to that of the polarization function in Eq. (\ref{Polarization b}) except the factor $F_{ss'}({\bf k, q})$ is replaced with $h_{ss'}({\bf k, q})$ given by,
\begin{align}
\label{Form factor Ha}
h_{ss'}({\bf k, q}) &= \frac{1}{4}\text {Tr}[(1+s' \hat \sigma_{\bm{k}}) \sigma^x (1 + s\hat \sigma_{\bm{k+q}})] \nonumber\\
				&=  \frac{1}{2}(s'\cos{\theta} + s\cos{\beta}).
\end{align}
where $\theta$ and $\beta$ are respectively the angles ${\bf k}$ and ${\bf k+q}$ makes with ${\bm q}$. Note that,
\begin{align}
\label{Form factor Hb}
h_{ss'}({\bf k, q}) = s'\cos{\theta} + s\frac{k\cos{\theta} + q}{|{\bf k+q}|}.
\end{align}
To calculate the integral, we use the same procedure as used above in Appendix A, taking the absolute values $k$ and $p =|{\bf k}+{\bf q}|$ as new variables, to replace the angle variable $\theta_{\bf k}$ in the integral over $d^2k =k \,dk \,d\theta_{\bf k}$ with  $p$.

\begin{align}
{\cal H}(\omega,q) & = 4 \int  \frac{dk dp }{(2\pi)^2} \frac{k p}{\sqrt{((k+p)^2 - q^2)(q^2-(k-p)^2)}}  \nonumber\\
 				& \times\frac{(n_{s'}^{\uparrow}({\cal E}_{\bm{k}}) - n_{s}^{\downarrow}({\cal E}_{\bm{p}}))}{\omega- {\cal E}^{s}_{\downarrow}(\bm{p})+{\cal E}^{s'}_{\uparrow}(\bm{k}) + i0^+} h_{ss'}({\bf k, q}),		
 \end{align}
 where the factor $h_{ss'}({\bf k, q})$ is,
\begin{eqnarray}
\label{Form factor Hc}
h_{ss'}({\bf k, q}) &=&  s\frac{(p-ss'k)}{q} F_{s,s'}({\bf k, q}).
\end{eqnarray}
From here, it follows easily to write the expression of ${\cal H}(\omega,q)$ using the same procedure as in Appendix A for polarization function for both intraband and interband transitions.
\subsection{Intraband transitions}
For intraband transitions, $s' = s$, we can write,
\begin{align}
\label{Intraband h1}
{\cal H}_{intra}(\omega,q) & = \frac{q^2}{8\pi^2}  \int\limits_{1}^{\infty}dx \sqrt{x^2-1} 
 				\int\limits_{-1}^{1} dy \frac{y}{\sqrt{1-y^2}} \nonumber\\
 				& \times \frac{\theta(2\omega_{B} - q(x-y))}{\Omega - qy + i0^+} .			
\end{align}
We are looking for the collective mode in the region where $|\Omega| < 0$ and $|\Omega| > q$, and, where the imaginary part of the above equation vanishes. For the real part we get,
\begin{equation}
\label{Intraband h2}
{\cal H}_{intra}(\omega, q) = \int\limits_{1}^{2\omega_{B}} \frac{dx}{8\pi^2}\sqrt{x^2 -q^2} \int\limits_{-1}^{1} \frac{dy~y}{\sqrt{1-y^2}(\Omega - qy)}.  
\end{equation} 
The $x-$integral gives a factor of $2\omega_B^2/q^2$. The $y-$ integral is calculated in the Appendix D and gives,
\begin{equation}
\label{Intraband h3}
{\cal H}_{intra}(\omega, q) = -\frac{\omega_{B}^2}{4\pi q}\bigg(1 - \frac{|\Omega|}{\sqrt{\Omega^2 - q^2}} \bigg).
\end{equation} 
\subsection{Interband transitions}
For interband transitions, $s' = -s$, we can write,
\begin{align}
\label{Interband h1}
{\cal H}_{inter} (\omega,& q) =  \frac{q^2}{16\pi^2}\int\limits_{1}^{\infty}dx \frac{x}{\sqrt{x^2-1}}\int\limits_{-1}^{1} dy\sqrt{1-y^2} \bigg(\frac{1}{\Omega - qx } \nonumber\\
			&-\!\frac{\theta(2\omega_{B} - q(x-y)) - \theta(q(x+y) - 2\omega_{B})}{\Omega + qx} \bigg).
\end{align} 
We now look at the real part of the polarization function coming from interband transitions given by the principal value of the integral in Eq. (\ref{Interband h1}).
For $\omega_{B} \gg q$, we can neglect $y$ in the two step functions in the second term since $y \in (-1,1)$. In doing so, $x$ and $y$ integrals becomes independent, the $y$ integral gives a factor of $\pi/2$ and the above equation becomes,
\begin{align}
\label{Interband re h1}
{\cal H}'_{inter}(\omega, q) = \frac{q^2}{16\pi} \bigg(\int\limits_{1}^{\infty} \frac{dx~\Omega x}{\sqrt{x^2 -1}(\Omega^2 - q^2 x^2)} \nonumber\\
										- \int\limits_{1}^{\frac{2\omega_{B}}{q} } \frac{dx~x}{\sqrt{x^2 -1}(\Omega+qx )}  \bigg).
\end{align} 
Note that the above integrals are taken in principal value sense, so that the first integral vanishes for $|\Omega|>q$. The second integral is calculated in Appendix D. For small q, we have,
\begin{align}
\label{Interband re h2}
{\cal H}'_{inter}(\omega, q) = -\frac{q}{16\pi}\bigg(\ln\frac{2\omega_{B}}{q} 
										+  \frac{\Omega}{\sqrt{\Omega^2 -q^2}} \ln\frac{2|\Omega|}{q} \bigg) .
\end{align} 
The imaginary part for $\Omega<0$ and $|\Omega| >q$ is given by,
\begin{align}
\label{Interband im h1}
{\cal H}''_{inter}(\omega, q) = -\frac{\Omega q}{16\pi\sqrt{\Omega^2 - q^2}} &
				 \int\limits_{-1}^{1} dy \sqrt{1-y^2}\big( \theta(qy+\omega ) \nonumber\\
				& - \theta(qy-\omega )\big).
\end{align}
The above integral has been calculated before (see Eq. (\ref{Interband polarization im a}) and the following section for discussion). 
In the region of interest, $\omega \gg q$,
\begin{align}
\label{Interband im h2}
{\cal H}''_{inter}(\omega, q) = -\frac{\Omega q}{32\sqrt{\Omega^2 -q^2}}.
\end{align} 
The function ${\cal H}(\omega,q) = {\mathcal H}_{intra}+ {\mathcal H}_{inter}$, combining the real and imaginary parts for intraband and interband transitions including the two Dirac points is,  
\begin{equation}
\label{Interband h2-appendix}
\! {\cal H}(\omega,q) = \frac{N\omega_{B}^2}{4\pi q} \bigg(\! -1 + \frac{|\Omega|}{\sqrt{\Omega^2 - q^2}} \bigg)
						       	 -\frac{iNq\Omega}{32\sqrt{\Omega^2 -q^2}}.							 					 
\end{equation}
Expanding ${\cal H}(\omega,q)$, near $q \approx 0$,
\begin{equation}
\label{Interband h3}
{\cal H}(\omega,q) = \frac{N\omega_{B}^2 q}{8\pi \Omega^2} .
\end{equation}

\section{Evaluation of Integrals}
i) The x-integral encountered in Eq. (\ref{Intraband polarization im a}) can be easily calculated by substituting, $x = \cosh t$,
\begin{equation}
\label{integral1}
I_1 = \int\limits_1^{\omega/q} dx \sqrt{x^2-1} = \int_0^b dt~ \sinh^2 t,
\end{equation}
where $b = \cosh^{-1}(\omega/q)$. The above integral is straightforward to calculate and gives,
\begin{align}
I_1 &= \frac{1}{4}(\sinh(2b) - 2b)  \nonumber\\
&= \frac{1}{2q^2} \Big(\omega \sqrt{\omega^2 - q^2 } -q^2\log\big(\frac{\omega + \sqrt{\omega^2 -q^2}}{q}\big)\Big).
\end{align}

ii) The y-integral in Eq. (\ref{Intraband polarization re b}) is calculated by substituting $y = \cos \phi$, giving us a simple known integral of the form,
\begin{equation}
\label{integral2}
I_2 = \int\limits_{-1}^{1} \frac{dy}{ \sqrt{1-y^2}} \frac{1}{|\Omega| + qy}  = \frac{1}{q}\int \limits_0^\pi \frac{d\phi}{a+ \cos \phi},
\end{equation}
where, $a = |\Omega/q| >1$. The above integral is a well-known integral, and utilize a contour integration in the complex plane. We integrate this over a unit circle in the complex plane, $z = e^{i \phi}$ centered at the origin,
\begin{equation}
I_2 =  \frac{-i}{q} \oint\, \frac{dz}{z^2+2az+1} = \frac{\pi}{q\sqrt{a^2-1}}.
\end{equation}

iii) The next y-integral to be calculated from Eq. (\ref{Current-current Intraband b}) can be solved again by substitution $y = \cos{\phi}$,
\begin{equation}
\label{integral3}
I_3 = \int\limits_{-1}^{1} dy \frac{y^2}{\sqrt{1-y^2}(\Omega +qy)},
\end{equation}
where, $a =  |\Omega/q| $, and with a little algebraic manipulation is related to the integral above ($a>1$) as,
\begin{align}
 I_3 = -&\frac{1}{q}\int \limits_0^\pi  \frac{d\phi~\cos^2{\phi}}{a - \cos \phi} = \frac{1}{q}\int \limits_0^\pi d\phi ~(a + \cos{\phi}) - a^2 I_2 \nonumber\\
       = & \frac{\pi}{q} \Big( a - \frac{a^2}{\sqrt{a^2-1}}\Big).
\end{align}

iv) Another y-integral that we need in Eq. (\ref{Intraband h2}) is solved similarly as above,
 \begin{equation}
 \label{integral4}
I_4 = \int\limits_{-1}^{1} dy \frac{y}{\sqrt{1-y^2}(\Omega -qy)}  = -\frac{\pi}{q} +\frac{|\Omega|}{q}I_2.
\end{equation}

v) The last integral to be calculated in Eq. (\ref{Interband re h1}) is,
\begin{align}
\label{integral5}
I_5 = &\int\limits_{1}^{\Lambda_B} \frac{dx~x}{\sqrt{x^2 -1}(\Omega+qx )} \nonumber\\
= & \frac{1}{q}\int\limits_{1}^{\Lambda_B} \frac{dx}{\sqrt{x^2 -1}} + \frac{1}{q}\int\limits_{1}^{\Lambda_B} \frac{dx~\Omega}{\sqrt{x^2 -1}(|\Omega|-qx)},
\end{align}
where $\Lambda_B = 2\omega_{B}/q$. The first integral is simply, $\ln \Lambda_B$. The second integral is solved by the substitution, $x = \cosh t = (e^t + e^{-t})/2$. And again, substituting, $z = e^t$ in the second integral,
\begin{equation}
I_5 = \frac{1}{q}\ln \Lambda_B  -  \frac{2\Omega}{q^2}\int\limits_{1}^{ 2\Lambda} dz\frac{1}{ (z - \Omega/q)^2 - (\Omega^2/q^2 -1)} .
\end{equation}
The above integral is a logarithmic function, where the upper limit can be extended to $\infty$ to give,
\begin{equation}
I_5 = \frac{1}{q}\Big(\ln \Lambda_B  -  \frac{\Omega}{\sqrt{\Omega^2-q^2}} \ln{\frac{\sqrt{\Omega^2 -q^2}-|\Omega|}{q}}\Big) .
\end{equation}

\section{Evaluations of the collective mode spectra}
In terms of the new dimensionless variables introduced as $z = |\Omega|/M$, and $q_0 = q/M $, the functions ${\cal P}$, $K$, and ${\cal H}$
\begin{subequations}
 \begin{align}
{\mathcal P}
 (z,q_{0}) = &-\frac{N}{u\sqrt{z^2-q_0^2}} - i\frac{M N q_0^2}{16\sqrt{z^2-q_0^2}}, \\
 {\cal K}
(z,q_{0}) = &-\frac{N z}{u q_0^2} \bigg(\!\! -1+ \frac{z}{\sqrt{z^2 - q_0^2}} \bigg)
									 - 2 N\tilde \Lambda \nonumber\\
						       	& -i\frac{NM z^2}{16\sqrt {z^2 -q_0^2}},\\
\! {\cal H}(z,q_{0}) = &\frac{N}{2u q_0} \bigg(\!\! -1+ \frac{z}{\sqrt{z^2 - q_0^2}}  \bigg)
						       	 +\frac{iNM q_0 z}{32\sqrt{z^2 -q_0^2}}.							 
 \end{align}
\end{subequations}
Substituting above functions in polarization function expression,  Eq. (\ref{Interacting polarization a}), and considering the case of weak interactions, $uN\tilde\Lambda\ll1$, we obtain
\begin{widetext}
\begin{eqnarray}
\Pi(z,q_0) &=& -\Big(\frac{2}{u}\Big) \frac{ q_0^2+ \sqrt{z^2-q_0^2}-z}{ \sqrt{z^2-q_0^2}(q_0^2-1+z) - (q_0^2-z+z^2) -i\frac{M u }{16}( q_0^2-z+z^2)q_0^2}. 
\end{eqnarray}

To study the collective mode dispersion for $q_0\ll z$, we have to expand the denominator of $\Pi(z,q_0)$ to the fourth order in $q_0$. At the same time, the leading term in the expansion of the numerator can be determined to the second order in $q_0$. As a result, we obtain a simplified expression for the full polarization function:
\begin{eqnarray}
\label{widetext pi}
\Pi(z,q_0)
&\approx& -\Big(\frac{2}{u}\Big) \frac{4z^2(2 z-1)}{4z^2(z-1)(2z-1) + q_0^2(1-z-4z^2) -i\frac{M u }{2}( q_0^2-z+z^2)z^3}. 
\end{eqnarray}
\end{widetext}

\bibliography{graphene-silin-refs.bib}
\bibliographystyle{apsrev}
\end{document}